\documentclass[twocolumn]{aastex6}
\usepackage{amsmath,color,textcomp,url,graphicx,subfigure}
\usepackage{multirow}
\newcolumntype{K}[1]{>{\centering\arraybackslash}p{#1}}

\def \MJ {$\,M_{\rm{Jup}}$}

\def \i {$i'$}
\def \z {$z'$}
\def \iz {$i'$ -- $z'$}
\def \BPMG {$\beta$PMG}
\def \ds {${d}_{s}$}
\def \dsp {${d}_{\rm{sp}}$}
\def \dph {${d}_{\rm{ph}}$}
\def \Ha {H$_{\alpha}$}
\def \Teff {$T_{\rm{eff}}$}
\def \logg {$\log g$}
\def \BANYAN {M13} %\citet{Malo2013} 
\def \BANYANIII {M14} %\citet{Malo2013}
\def \BANYANII {G14} %\citet{Gagne2014} 
\def \plmo {\,$\pm$\,}
\def \env {$\sim$\,}
\def \gta {\,\lower 0.5ex\hbox{$\buildrel > \over \sim\ $}\,}   %greater than about
\def \lta {\,\lower 0.5ex\hbox{$\buildrel < \over \sim\ $}\,}   %less than about

\usepackage{soul} % pour pouvoir barrer du texte
\usepackage{color}
\definecolor{orange}{rgb}{0.9,0.6,0.2} 

\begin{document}
\title{PSYM-WIDE: A Survey for Large-separation Planetary-mass Companions to Late Spectral Type Members of Young Moving Groups}
\author{Marie-Eve Naud\altaffilmark{1}*, \'{E}tienne Artigau\altaffilmark{1}, Ren\'{e} Doyon\altaffilmark{1}, Lison Malo\altaffilmark{2,1},  Jonathan Gagn\'{e}\altaffilmark{3,4}, David Lafreni\`{e}re\altaffilmark{1}, Christian Wolf\altaffilmark{5} and Eugene A. Magnier\altaffilmark{6}} 
\email{*Corresponding author: naud@astro.umontreal.ca}
\affil{\altaffilmark{1} Institut de recherche sur les exoplan\`{e}tes, D\'{e}partement de physique, Universit\'{e} de Montr\'{e}al, Montr\'{e}al, QC H3C 3J7, Canada.\\
\altaffilmark{2} Canada-France-Hawaii Telescope (CFHT) Corporation, 65-1238 Mamalahoa Highway, Kamuela, HI 96743, USA.\\
\altaffilmark{3} Department of Terrestrial Magnetism, Carnegie Institution for Science, 5241 Broad Branch Road NW, Washington, DC 20015, USA.\\
\altaffilmark{4} NASA Sagan Fellow\\
\altaffilmark{5} Research School of Astronomy and Astrophysics, Australian National University, Canberra, ACT 2611, Australia\\
\altaffilmark{6} Institute for Astronomy, University of Hawaii, 2680 Woodlawn Drive, Honolulu, HI 96822, USA.}

%%%%%%%%%%%%%%%%%%%%%%%%%%%%%%%%%%%%%%%%%%%%%%%%%%%%%%%%%%%%%%%%%%%%%%%%%
\begin{abstract}
We present the results of a direct-imaging survey for very large separation ($>$100\,au), companions around  95 nearby young K5--L5 stars and brown dwarfs. They are high-likelihood candidates or confirmed members of the young ($\lta$150\,Myr) $\beta$ Pictoris and AB Doradus moving groups (ABDMG) and the TW Hya, Tucana-Horologium, Columba, Carina, and Argus associations. Images in \i\ and \z\ filters were obtained with the Gemini Multi-Object Spectrograph (GMOS)  on Gemini South to search for companions down to an apparent magnitude of \z\env22--24 at separations \gta20\arcsec\ from the targets and in the remainder of the wide 5$\farcm$5 $\times$ 5$\farcm$5 GMOS field of view. This allowed us to probe the most distant region where planetary-mass companions could be gravitationally bound to the targets. This region was left largely unstudied by past high-contrast imaging surveys, which probed much closer-in separations. This survey led to the discovery of a planetary-mass (9--13\,\MJ) companion at 2000\,au from the M3V star GU Psc, a highly probable member of ABDMG. No other substellar companions were identified. These results allowed us to constrain the frequency of distant planetary-mass companions (5--13\,\MJ) to 0.84$_{-0.66}^{+6.73}$\% (95\% confidence) at semimajor axes between 500 and 5000\,au around young K5--L5 stars and brown dwarfs. This is consistent with other studies suggesting that gravitationally bound planetary-mass companions at wide separations from low-mass stars are relatively rare.

\end{abstract}

%%%%%%%%%%%%%%%%%%%%%%%%%%%%%%%%%%%%%%%%%%%%%%%%%%%%%%%%%%%%%%%%%%%%%%%%%
\section{Introduction}
\label{introduction}
Twenty years after the first detection of an exoplanet around a main-sequence star \citep{Mayor1995}, the increasing number of known exoplanets provides a clearer overall picture of the content and architecture of exoplanetary systems. 
However, the outer realms of planetary systems, inaccessible to the radial velocity and transit methods, are still largely unexplored. 
Direct imaging is the prime method for exploring separations larger than a few tens of astronomical units. This method has seen tremendous improvements since the first major discoveries, including the first image of a planetary-mass companion around the brown dwarf 2MASS~J12073346-3932539~b (2M~1207~b hereafter; \citealp{Gizis2002,Chauvin2004,Ducourant2008}), the first image of a planet around a sun-like star, 1RXS~J1609-2105~b \citep{Lafreniere2008,Lafreniere2010} and the first exoplanetary system, around HR~8799 \citep{Marois2008,Marois2010}. Dedicated second-generation, high-contrast imagers like SPHERE \citep{Beuzit2008} and GPI \citep{Macintosh2014} are now reaching contrasts allowing the detection of giant planets from \env5 to \env100\,au \citep{Macintosh2015,Wagner2016}. 

While similar to their closer-in exoplanet counterparts in many ways, distant, directly imaged companions also share similarities with low-mass brown dwarf companions and isolated planetary-mass objects (e.g., \citealp{Faherty2016}). The directly imaged exoplanets found to date provide essential constraints on the dynamics of planetary systems and on substellar formation models and come with their own open questions. Most of them are not readily explained by standard planetary formation scenarios. They could be planets formed in a disk that were later scattered outward or planetary-mass objects that formed like brown dwarfs and stars, through the fragmentation of a collapsing prestellar core. 

Young stars are prime targets for direct imaging surveys, as young companions are brighter than their older counterparts, since they are still contracting and cooling down. Recently, significant progress has been made to identify young stars of the local neighborhood that are members of Young Moving Groups (YMGs). Stars in these sparse ensembles were formed together and therefore share similar positions and space motions in the Galaxy \citep{ZuckermanSong2004}. Their members provide an important advantage for direct imaging surveys, because evolutionary models allow us to translate their well-constrained age to relatively precise mass constraints for planetary-mass companions. Most low-mass late spectral type members of these associations remained undetected until a few years ago because the observations used to determine proper motions, radial velocities, and distances were mostly available in the optical. \citeauthor{Malo2013} (\citeyear{Malo2013}; M13 hereafter), \citeauthor{Malo2014b} (\citeyear{Malo2014b}; M14 hereafter), and \citeauthor{Gagne2014a} (\citeyear{Gagne2014a}; G14 hereafter) identified a large number of low-mass stars, brown dwarfs, and isolated planetary-mass objects with high membership probabilities in seven young and nearby YMGs (the $\beta$ Pictoris moving group, $\beta$PMG; the TW Hya association, TWA; the Tucana-Horologium association, THA; the Columba association, COL; the Carina association, CAR; the Argus association, ARG; and the AB Doradus moving group, ABDMG), using a novel Bayesian analysis and dedicated observation programs. 

Some of the first direct imaging surveys concentrated on massive stars, where theory predicts more giant exoplanets and where some of the first detections of planets through direct imaging were made (notably, HR 8799, an A5V star; \citealp{Marois2008,Marois2010}). First-generation surveys, like the Gemini Deep Planet Survey (GDPS; \citealp{Lafreniere2007}) and the NaCo Deep imaging survey of young, nearby austral stars \citep{Chauvin2010} did include several M stars. Interestingly, the latter led to the discovery of the planetary-mass companion around the M8 brown dwarf 2M1207. Surveys dedicated to low-mass stars were undertaken in recent years. The PALMS survey (Planets Around Low-Mass Stars; \citealp{Bowler2015}) did not detect any 1--13\,\MJ\ companions between 10--100\,au around their sample of 122 K5--M4 single dwarfs. This allowed determination of an upper limit (95\% confidence level) of 10.3\% (16\%) for these objects, assuming a hot (cold) start evolutionary model. \citet{Lannier2016} presents the resulta of another M-star survey, based on VLT observations. Their sample of 58 M stars includes most of the 16 stars from the \citet{Delorme2012a} survey, a pioneer study dedicated to low-mass stars. A frequency of $2.3_{ -0.7}^{+2.9}\%$ is determined for 2--14\,\MJ\ companions at separations of 8--400\,au. The meta-analysis presented by \citet{Bowler2016}, which summarizes the results of nine surveys (including PALMS, GDPS, and the Gemini NICI Planet-Finding Campaign; \citealp{Biller2013}), includes 118 M stars and finds an upper limit of 3.9\% (5.4\%; 7.3\%) for the occurrence of 5--13\,\MJ\ at 30--300\,au (10--1000\,au; 100--1000\,au) around them. The results of the IDPS (International Deep Planet Search) survey (292 stars) were combined with those of GDPS and of the NaCo-LP survey \citep{Chauvin2015} in \citet{Galicher2016}. They find a planetary-mass (0.3--14\,\MJ) companion fraction between 20--300\,au of $0.90_{-0.65}^{+4.05}\%$ for their ``low mass'' ($<$1.1\,$M_{\odot}$) sample, which includes G, K, and M stars.

In 2010, the survey PSYM -- Planet Search around Young-associations M dwarfs -- was started to detect planetary-mass companions around young K5--L5 stars and brown dwarfs newly identified in M13, M14, and G14. This paper presents the results of the PSYM-WIDE survey of 95 stars with the Gemini Multi-Object Spectrograph (GMOS; \citealp{Hook2004}) at Gemini South. PSYM-WIDE was designed specifically to detect planetary-mass companions at large (500--5000\,au) separations. A new planetary-mass companion, GU~Psc~b, was identified as part of this survey and was presented by \citet{Naud2014}. The sample and selection criteria are described in \S 2, and the observations are presented in \S 3, and followed by the results in \S 4. A discussion that puts the results derived in perspective is presented in \S 5. The paper concludes with a discussion on the plausible origin of these wide companions and ongoing efforts to find them.

%%%%%%%%%%%%%%%%%%%%%%%%%%%%%%%%%%%%%%%%%%%%%%%%%%%%%%%%%%%%%%%%%%%%%%%%%
\section{The stellar sample}
\label{the-star-sample}

\subsection{Target Selection}
\label{subsec:target}
The sample of stars surveyed in this work has been drawn primarily from high-probability YMG members identified by the Bayesian analysis presented in M13, M14, and G14. The BANYAN (\BANYAN, \BANYANIII) and BANYAN~II (\BANYANII) tools both use sky position, proper motion, asnd color-magnitude diagrams to assess the probability that a star is a member of \BPMG, ABDMG, TWA, THA, COL, CAR, or ARG. The Bayesian analysis provides an estimation of the radial velocity and distance (statistical distance; \ds) of a star assuming membership to a given association. The statistical distance and predicted radial velocities have been demonstrated to have a typical accuracy of \env10--20\% compared to direct measurements when membership is confirmed (see \BANYAN). When a star has a high membership probability, this method therefore provides good estimates of those values. Measuring the radial velocity or parallax together with other signs of youth is needed to unambiguously establish the membership of a candidate member. 

In \BANYAN, the $I_{C}$ and $J$ photometry was used with the BANYAN tool to identify 214 new, highly probable low-mass members (spectral types K5--M5) among an initial sample of several hundreds of stars displaying youth indicators such as \Ha\ or X-ray emission from \citet{Riaz2006}. In \BANYANIII, new radial velocity measurements were included in the analysis to further confirm the membership of 130 candidates from \BANYAN\ and 57 other stars from the literature. The BANYAN~II tool presented in \BANYANII\ adapted the \BANYAN\ analysis to identify lower-mass stars and brown dwarf (later than M7) members of the YMG, using 2MASS and $WISE$ photometry. Their initial candidate sample is composed of 158 stars that display spectroscopic signs of youth or have unusually red colors for their spectral type at near-infrared wavelengths. Among these, 25 new high-probability candidates were identified, and the membership of 10 candidates was confirmed. The same tool was used in an all-sky survey built from a cross-match of the 2MASS and AllWISE to identify a total of 228 new M4--L6 candidate members of YMGs \citep{Gagne2015a, Gagne2015c}.

Among the \BANYAN/\BANYANIII/\BANYANII\ published or preliminary samples, those with declinations lower than $+$20$^{\circ}$ were first selected, as observations were to be made at Gemini South in Chile. Stars with the highest membership probabilities were prioritized. Stars in the youngest associations were preferred, as younger companions at a given mass are brighter than their older counterparts and thus easier to detect. Stars with the nearest statistical distances (or parallaxes when available) were also prioritized, in order to probe a region as close as possible to the stars. Objects located at distances beyond 80\,pc were rejected. 
Binary stars were not excluded a priori from the selection. Twenty stars in the sample are known as double or triple systems. These are identified in the spectral type column of Table~1 with the mention ``sb1'', ``sb2'' or ``sb3'', or with the ``+'' sign, which indicates that there is a stellar companion (the spectral type of this companion is sometimes not known). Recent discoveries have demonstrated that the presence of a similar-mass or lower-mass companion does not preclude the detection of additional companions around a star; Ross 458(AB)c represents such a low-mass companion on a very wide orbit around a much tighter M-dwarf binary \citep{Goldman2010}. A total of 69 stars were taken from the \BANYAN/\BANYANIII\ sample, and 12 from \BANYANII.

Seven bona fide members previously known in the literature and used in \BANYAN\ or \BANYANII\ to determine the kinematic and photometric properties of the YMGs were also added to the sample. A few young stars that do not appear in M13, M14, or G14 but that were also identified as young in the literature (three from \citealp{Shkolnik2011,Shkolnik2012}, three from \citealp{Rodriguez2011}, and one from \citealp{Kiss2011}) were also included. 

\begin{figure}[htbp]
\begin{center}
\includegraphics[width=8.5cm]{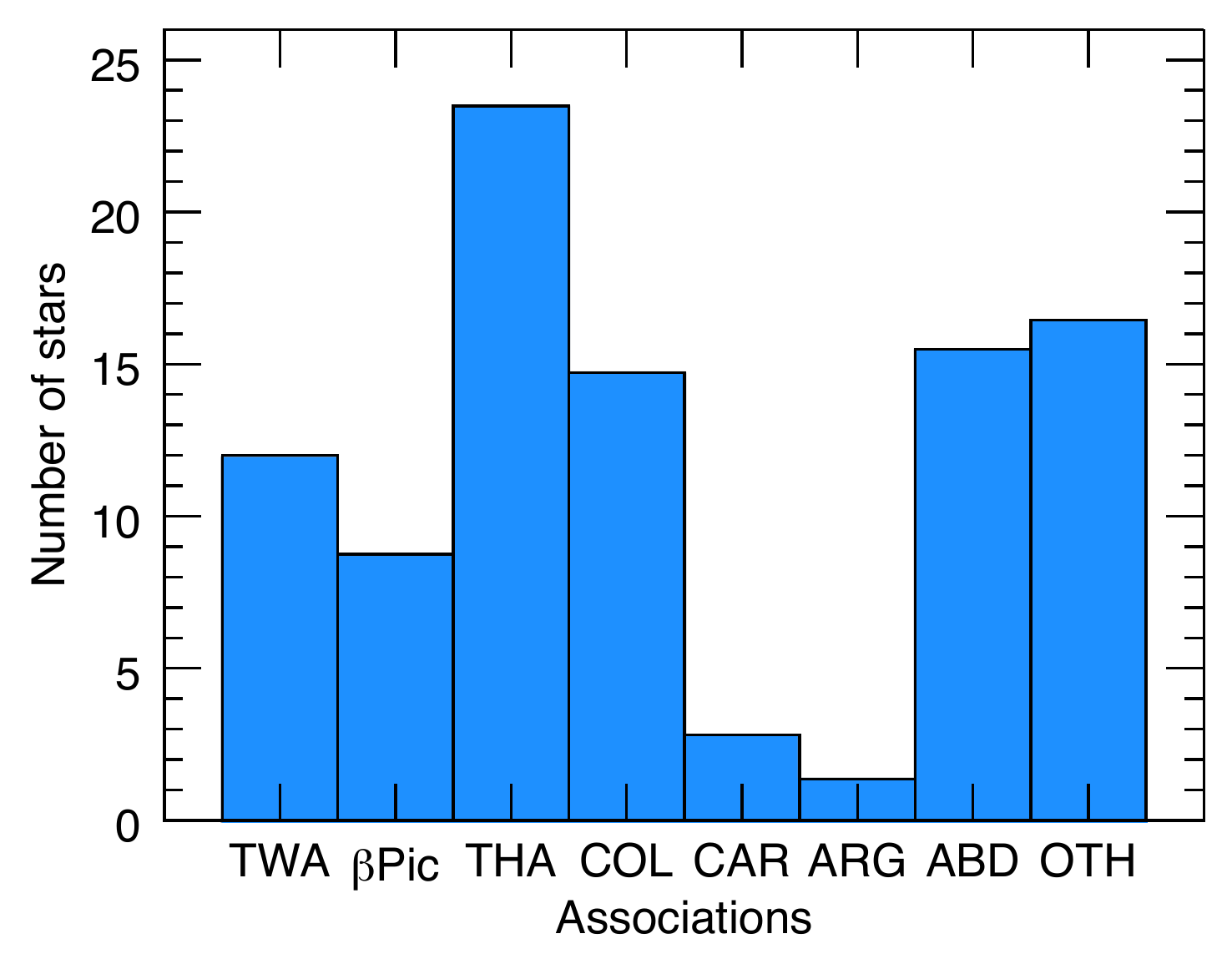}
\caption{Distribution of the target stars most probable associations of the target stars.}
\label{fig:star-prop-histo_1}
\end{center}
\end{figure}

\begin{figure*}[htbp]
\begin{center}
\includegraphics[width=15cm]{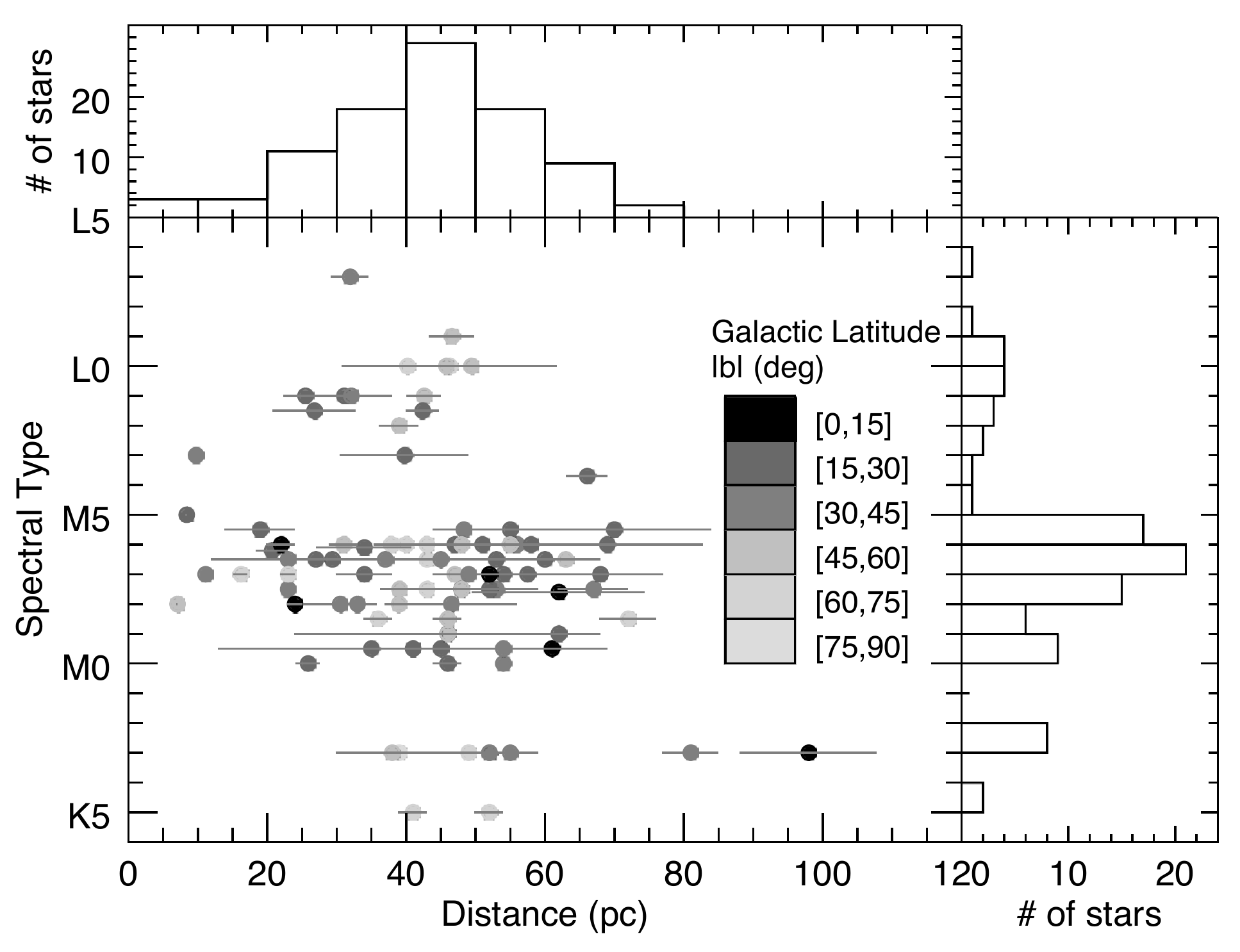}
\caption{Distribution of spectral types vs. distribution of distances. The histograms of these values are also shown. The galactic latitude $|b|$ is color coded: the greater the distance from the galactic plane, the lighter the points are.}
\label{fig:star-prop-histo_2}
\end{center}
\end{figure*}

The properties of the final sample of 95 stars are listed in Table~1 and presented in Figures \ref{fig:star-prop-histo_1} and \ref{fig:star-prop-histo_2}. They have late spectral types ranging from K5 to L5, with a median type of M3. The least massive of the stars in the sample are close to the deuterium-burning limit mass. For example, \citet{Faherty2016} estimated the mass of the L3 2MASS J21265040-8140293 to be 24.21$\pm$14.3\,\MJ, and that of the L1 2MASS J00040288-6410358 to be 16.11$\pm$2.9\,\MJ. No selection was made based on the galactic latitude; seven targets have galactic latitude $|b|<15^{\circ}$, and are thus located in relatively crowded fields. This slightly complicates the confirmation procedure and reduces the likelihood of planet detection (see Section \ref{contrast}). It is important to note that the sample of young nearby stars from which we draw our sample is still under construction and suffers many biases (\citealt{Riaz2006}, for example, only selected the sources that are bright in X-ray). Therefore, it is not expected that it follows closely a field initial mass function.

\subsection{Age and Distance estimates}
\label{subsec:the-star-sample}

A distance estimate for the target star is needed to convert angular separation to physical separation and apparent magnitude limits to absolute magnitude limits. An estimate of the age is also necessary to convert absolute magnitude to mass, using evolutionary models. Assigning membership to a young association is one of the few ways that are available to constrain the age of low-mass stars and obtain an approximation of their distance, as seen in Section \ref{subsec:target}. All targets selected for the PSYM-WIDE survey were analyzed with the most recent version of BANYAN (spectral type earlier than M7) or BANYAN II (spectral type later than M7) to calculate their membership probability to several YGMs, informed by the most recent measurements of proper motion, parallax, and radial velocity. The membership of all stars is listed in Table~2. 

The status ``bona fide'' (BF) was assigned to stars with all kinematic measurements, a trigonometric parallax, and youth indicators that have a Bayesian probability above a selected high threshold ($>$90\% for stars analyzed with BANYAN and $>99\%$ for those analyzed with BANYAN~II) that minimizes the chance of a false positive in the sample. Objects that are missing one kinematic measurement and have a Bayesian probability above the threshold are referred to as ``high-likelihood candidates'' (HLC). Those that have no radial velocity or parallax measurements with a Bayesian probability above the threshold are referred to as ``candidates'' (C). The large majority of the stars in the sample belong to one of these categories (7 BF, 58 HLC, and 2 C). 

Ten stars have an ambiguous membership status (AY for ``ambiguous membership, young''), because their membership probability is high in two or more of the seven associations. Seventeen stars were assigned the status ``young other'' (YO). Such cases correspond to  stars for which the BANYAN membership probability assigned is low but nonnegligible for at least one moving group, members of YMGs that are not known or not included in BANYAN, or simply relatively young stars that do not belong to a group. In one case, a star initially thought young was found to display no youth indicator. It has the status NYI (``no youth indicator'') in Table~2. 

The histogram of Figure~\ref{fig:star-prop-histo_1} shows the most probable association for all stars. Candidate members of TWA, \BPMG, THA, and COL are the most numerous as they are the youngest associations and were thus favored in the sample construction. Several stars are also candidate members of ABDMG.

\subsubsection{Age}
For BF, HLC, C, and AY stars, the total age range of all the plausible association(s) is conservatively assigned to the star. The association age ranges determined in the recent analysis of \citet{Bell2015} are used here: \BPMG: 24\plmo 3\,Myr; ABDMG: $149^{+51}_{-19}$\,Myr; TWA: 10\plmo3\,Myr; THA: 45\plmo4\,Myr; COL: $42^{+5}_{-4}$\,Myr; 
\clearpage
%dans cette version de la table, j'ai complètement réécris les notes. J'ai aussi enlevé toutes les références ``v`'' (Riaz2006) pour les mettre a la fin. Faire la meme chose avec (ii)  Zacharias et al. 2012 pour les references mag I
%Corrigé J01415823-4633574 pour chiffre significatif du proper motion: -49$\pm$10
% ajouté \renewcommand\thetable{1}
% Enlevé les ^{\rm{s}} (c'est considéré dans la légende, quand on dit que si on dit rien, c'est ``s``''''
% Mis en premier les ``unless otherwise staten'' dans la note en bas de page ``b''
%\rm{a} pour les tablenote en bas de la table (pour pas que ce soit en italique)
% Ajusté les chiffres significatifs des trigo. distance à la mitaine
% ajouté une cline en dessous de J, H, K
\floattable
\begin{deluxetable}{hchcccccccccccccc}%[htbp]
\renewcommand\thetable{1}
\renewcommand{\tabcolsep}{1.5mm}
\tablewidth{650pt}
\tabletypesize{\footnotesize}
\rotate
\tablecaption{Target Sample Properties}
\label{tab-stardata}
\tablehead{
\nocolhead{Index}&\multicolumn{2}{c}{2MASS Designation}&\multicolumn{2}{c}{Coordinates}&\multicolumn{3}{c}{Proper Motion}&\colhead{Sp.Type\tablenotemark{a}$^{,}$\tablenotemark{b}}&\multicolumn{6}{c}{Magnitudes}&\colhead{Trigonometric}&\colhead{Radial Velocity\tablenotemark{b}}\\
\cline{6-8}
\cline{10-15}
\nocolhead{}&\colhead{}&\nocolhead{Other}&\colhead{$\alpha$}&\colhead{$\delta$}&\colhead{$\mu_{\alpha}\cos{\delta}$}&\colhead{$\mu_{\delta}$}&\colhead{Ref.\tablenotemark{b}}&\colhead{(Opt.)}&\colhead{$I$\tablenotemark{b}}&\colhead{$J$}&\colhead{$H$}&\colhead{$K_S$}&\colhead{$W1$}&\colhead{$W2$}&\colhead{distance\tablenotemark{b}}&\colhead{}\\
\cline{11-13}
\nocolhead{}&\colhead{}&\nocolhead{}&\colhead{(J2000.0)}&\colhead{(J2000.0)}&\colhead{(mas yr$^{-1}$)}&\colhead{(mas yr$^{-1}$)}&\colhead{}&\colhead{}&\colhead{}&\multicolumn{3}{c}{(2MASS)}&\colhead{}&\colhead{}&\colhead{(pc)}&\colhead{(km s$^{-1}$)}
}
\startdata
0&J00040288-6410358&&   1.0120& -64.1766&  64.0$\pm$12.0& -47.0$\pm$12.0&F16&L1 $\gamma$$^{\rm{o}}$&& 15.79& 14.83& 14.01& 13.41& 12.96&&$5.3\pm3.4^{\rm{l}}$\\
1&J00172353-6645124&&   4.3481& -66.7535& 102.9$\pm$ 1.0& -15.0$\pm$ 1.0&Z12&M2.5&$10.66$&  8.56&  7.93&  7.70&  7.59&  7.50&$39.1\pm2.6^{\rm{y}}$&$10.7\pm0.2$\\
2&J00325584-4405058&&   8.2327& -44.0850& 128.3$\pm$ 3.4& -93.6$\pm$ 3.0&F16&L0 $\gamma$$^{\rm{g}}$&& 14.78& 13.86& 13.27& 12.84& 12.52&$46.3\pm15.4^{\rm{k}}$&$12.9\pm1.9^{\rm{k}}$\\
3&J00374306-5846229&&   9.4294& -58.7730&  57.0$\pm$10.0&  17.0$\pm$ 5.0&F16&L0 $\gamma$$^{\rm{g}}$&& 15.37& 14.26& 13.59& 13.15& 12.77&&$6.6\pm0.1^{\rm{k}}$\\
4&J01071194-1935359&&  16.7998& -19.5933&  64.4$\pm$ 1.6& -39.5$\pm$ 1.2&Z12&M0.5+M2.5$^{\rm{c}}$&$9.42^{\rm{i}}$&  8.15&  7.47&  7.25&  7.09&  7.11&&$11.5\pm1.4^{\rm{p}}$\\
5&J01123504+1703557&&  18.1460&  17.0655&  92.0$\pm$ 1.0& -98.4$\pm$ 1.0&Z05&M3&$12.23$& 10.21&  9.60&  9.35&  9.26&  9.13&&$-1.5\pm0.5$\\
6&J01132958-0738088&&  18.3733&  -7.6358&  70.5$\pm$ 1.1& -66.1$\pm$ 1.0&Z12&K7+M5.5$^{\rm{n}}$&$10.95$&  9.36&  8.71&  8.53&  8.43&  8.41&&$41.3\pm4.1$\\
7&J01220441-3337036&&  20.5184& -33.6177& 105.3$\pm$ 1.2& -58.3$\pm$ 1.0&Z12&K7&$9.92$&  8.31&  7.64&  7.45&  7.27&  7.37&&$4.7\pm0.4$\\
8&J01351393-0712517&&  23.8080&  -7.2144& 106.5$\pm$ 5.1& -60.7$\pm$ 5.1&Ro10&M4(sb2)$^{\rm{v,s}}$&$10.52^{\rm{i}}$&  8.96&  8.39&  8.08&  7.97&  7.80&$37.9\pm2.4^{\rm{dd}}$&$6.8\pm0.8$\\
9&J01415823-4633574&&  25.4926& -46.5660& 105.0$\pm$10.0& -49$\pm$10&F16&L0 $\gamma$$^{\rm{g}}$&& 14.83& 13.88& 13.10& 12.58& 12.19&&$6.4\pm1.6^{\rm{k}}$\\
10&J01484087-4830519&&  27.1703& -48.5144& 110.3$\pm$ 1.1& -51.0$\pm$ 1.1&Z12&M1.5&$11.04$&  9.19&  8.55&  8.36&  8.26&  8.19&&$21.5\pm0.2$\\
11&J01521830-5950168&&  28.0763& -59.8380& 109.2$\pm$ 1.8& -25.7$\pm$ 1.8&Z12&M2-3$^{\rm{p}}$&$10.83$&  8.94&  8.33&  8.14&  7.96&  7.88&&$8.1\pm1.8$\\
12&J02045317-5346162&&  31.2216& -53.7712&  95.1$\pm$ 2.9& -33.6$\pm$ 3.1&Z12&K5&$12.85$& 10.44&  9.81&  9.56&  9.41&  9.22&&$10.9\pm0.3$\\
13&J02070176-4406380&&  31.7573& -44.1106&  94.9$\pm$ 1.3& -30.6$\pm$ 1.3&Z12&M3.5(sb1)$^{\rm{v,s}}$&$11.28$&  9.27&  8.69&  8.40&  8.25&  8.09&&$10.1\pm0.3$\\
14&J02155892-0929121&&  33.9955&  -9.4867&  96.6$\pm$ 1.9& -46.5$\pm$ 2.6&Z12&M2.5(sb3)$^{\rm{v,s}}$&$9.79^{\rm{i}}$&  8.43&  7.80&  7.55&  7.31&  7.26&&$2.5\pm0.3$\\
15&J02215494-5412054&&  35.4790& -54.2015& 136.0$\pm$10.0& -10.0$\pm$17.0&F16&M8 $\beta$$^{\rm{u}}$&& 13.90& 13.22& 12.66& 12.34& 11.97&&$10.2\pm0.1^{\rm{k}}$\\
16&J02224418-6022476&&  35.6841& -60.3799& 137.4$\pm$ 1.7& -13.8$\pm$ 1.7&Z12&M4&$11.24$&  8.99&  8.39&  8.10&  7.95&  7.80&&$13.1\pm0.9$\\
17&J02251947-5837295&&  36.3311& -58.6249& 102.2$\pm$ 5.2& -25.0$\pm$ 7.3&2MAW&M9 $\beta$$^{\rm{k}}$&& 13.74& 13.06& 12.56& 12.26& 11.96&&\\
18&J02303239-4342232&&  37.6350& -43.7065&  80.3$\pm$ 0.9& -13.3$\pm$ 0.9&Z12&K5Ve*$^{\rm{ee}}$&$9.36$&  8.02&  7.43&  7.23&  7.12&  7.22&&$16.0\pm1.3$\\
19&J02340093-6442068&&  38.5039& -64.7019&  88.0$\pm$12.0& -15.0$\pm$12.0&F16&L0 $\gamma$$^{\rm{o}}$&& 15.32& 14.44& 13.85& 13.27& 12.93&&$11.8\pm0.7^{\rm{k}}$\\
20&J02485260-3404246&&  42.2192& -34.0735&  90.2$\pm$ 1.4& -23.7$\pm$ 1.4&Z12&M4(sb1)$^{\rm{v,s}}$&$13.64$&  9.31&  8.63&  8.40&  8.25&  8.05&&$14.6\pm0.3$\\
21&J02564708-6343027&&  44.1962& -63.7174&  67.4$\pm$ 2.2&   8.3$\pm$ 5.6&Z12&M4&$11.31^{\rm{i}}$&  9.86&  9.22&  9.01&  8.80&  8.63&&$18.5\pm3.4$\\
22&J03050976-3725058&&  46.2907& -37.4183&  50.8$\pm$ 1.3& -12.2$\pm$ 1.3&Z12&M1.5+M3$^{\rm{c}}$&$11.46$&  9.54&  8.88&  8.65&  8.56&  8.46&&$14.3\pm0.6$\\
23&J03350208+2342356&&  53.7587&  23.7099&  54.0$\pm$10.0& -56.0$\pm$10.0&F16&M8.5$^{\rm{t}}$&& 12.25& 11.65& 11.26& 11.06& 10.77&$42.4\pm2.3^{\rm{dd}}$&$15.5\pm1.7^{\rm{dd}}$\\
24&J03494535-6730350&&  57.4390& -67.5097&  41.8$\pm$ 1.0&  20.5$\pm$ 1.0&Z12&K7&$11.16$&  9.85&  9.23&  9.03&  8.87&  8.88&&$16.8\pm0.2$\\
25&J04082685-7844471&&  62.1119& -78.7464&  54.7$\pm$ 1.4&  42.1$\pm$ 1.4&Z12&M0&$10.89$&  9.28&  8.59&  8.40&  8.29&  8.26&&$16.4\pm0.4$\\
26&J04091413-4008019&&  62.3089& -40.1339&  45.9$\pm$ 1.7&   7.2$\pm$ 1.7&Z12&M3.5&$12.82$& 10.65& 10.00&  9.77&  9.68&  9.52&&$21.3\pm0.5$\\
27&J04213904-7233562&&  65.4127& -72.5656&  62.2$\pm$ 1.3&  26.6$\pm$ 1.3&Z12&M2.5&$11.82$&  9.87&  9.25&  8.99&  8.91&  8.79&&$15.0\pm0.3$\\
28&J04240094-5512223&&  66.0040& -55.2062&  42.4$\pm$ 2.1&  17.2$\pm$ 2.1&Z12&M2.5&$11.75$&  9.80&  9.16&  8.95&  8.80&  8.67&&$20.1\pm0.5$\\
29&J04363294-7851021&&  69.1373& -78.8506&  33.0$\pm$ 3.0&  47.0$\pm$ 2.7&Z12&M4&$12.52^{\rm{i}}$& 10.98& 10.36& 10.10&  9.96&  9.77&&$26.5\pm0.3$\\
30&J04365738-1613065&&  69.2391& -16.2185& 109.8$\pm$ 3.0& -21.9$\pm$ 4.2&Z12&M3.5&$11.30$&  9.12&  8.47&  8.26&  8.14&  7.98&&$15.7\pm0.5$\\
31&J04402325-0530082&LP 655-48&  70.0969&  -5.5023& 320.4$\pm$10.6& 126.8$\pm$ 7.3&2MAW&M7$^{\rm{e}}$&& 10.66&  9.99&  9.55&  9.36&  9.17&$9.8\pm0.1^{\rm{y}}$&$29.9\pm0.2^{\rm{dd}}$\\
32&J04433761+0002051&2MUCD 10320&  70.9067&   0.0348&  28.0$\pm$14.0& -99.0$\pm$14.0&F16&M9 $\gamma$$^{\rm{f}}$&& 12.51& 11.80& 11.22& 10.83& 10.48&&$17.0\pm0.8^{\rm{k}}$\\
33&J04440099-6624036&&  71.0041& -66.4010&  51.6$\pm$ 2.6&  33.3$\pm$ 2.6&Z12&M0.5&$11.05$&  9.47&  8.75&  8.58&  8.50&  8.47&&$16.7\pm0.4$\\
34&J04480066-5041255&&  72.0028& -50.6904&  53.1$\pm$ 2.1&  15.7$\pm$ 2.3&Z12&K7&$10.42$&  8.74&  8.08&  7.92&  7.81&  7.79&&$19.3\pm0.1$\\
\enddata
\end{deluxetable}
\floattable
\begin{deluxetable}{hchcccccccccccccc}%[htbp]
\renewcommand\thetable{1}
\tabletypesize{\footnotesize}
\rotate
\tablecaption{\textit{continued}}
\label{tab:stardata}
\tablehead{
\nocolhead{Index}&\multicolumn{2}{c}{2MASS Designation}&\multicolumn{2}{c}{Coordinates}&\multicolumn{3}{c}{Proper Motion}&\colhead{Sp.Type\tablenotemark{a}$^{,}$\tablenotemark{b}}&\multicolumn{6}{c}{Magnitudes}&\colhead{Triginometric}&\colhead{Radial Velocity\tablenotemark{b}}\\
\cline{6-8}
\cline{10-15}
\nocolhead{}&\colhead{}&\nocolhead{Other}&\colhead{$\alpha$}&\colhead{$\delta$}&\colhead{$\mu_{\alpha}\cos{\delta}$}&\colhead{$\mu_{\delta}$}&\colhead{Ref.\tablenotemark{b}}&\colhead{(Opt.)}&\colhead{$I$\tablenotemark{b}}&\colhead{$J$}&\colhead{$H$}&\colhead{$K_S$}&\colhead{$W1$}&\colhead{$W2$}&\colhead{distance\tablenotemark{b}}&\colhead{}\\
\cline{11-13}
\nocolhead{}&\colhead{}&\nocolhead{}&\colhead{(J2000.0)}&\colhead{(J2000.0)}&\colhead{(mas yr$^{-1}$)}&\colhead{(mas yr$^{-1}$)}&\colhead{}&\colhead{}&\colhead{}&\multicolumn{3}{c}{(2MASS)}&\colhead{}&\colhead{}&\colhead{(pc)}&\colhead{(km s$^{-1}$)}
}
\startdata
35&J04533054-5551318&HIP 22738 AB&  73.3773& -55.8588& 134.5$\pm$ 2.4&  72.7$\pm$ 2.0&vL07&M3Ve+M3Ve$^{\rm{r}}$&$8.15^{\rm{q}}$&  7.80&  7.24&  6.89&  5.96&  5.38&$11.1\pm0.2^{\rm{ff}}$&$30.0\pm0.0^{\rm{ee}}$\\
36&J04571728-0621564&&  74.3220&  -6.3657&  22.9$\pm$ 1.9& -99.1$\pm$ 2.5&Z12&M0.5&$11.11$&  9.51&  8.83&  8.64&  8.53&  8.51&&$23.4\pm0.3$\\
37&J04593483+0147007&HIP 23200&  74.8951&   1.7835&  34.6$\pm$ 2.3& -94.3$\pm$ 1.4&vL07&M0Ve$^{\rm{r}}$&$8.21^{\rm{a}}$&  7.12&  6.45&  6.26&  6.21&  6.06&$25.9\pm1.7^{\rm{ff}}$&$19.8\pm0.0^{\rm{b}}$\\
38&J05090356-4209199&&  77.2649& -42.1555&  26.7$\pm$ 1.8&  59.0$\pm$ 1.4&Z12&M3.5&$11.72$&  9.58&  8.98&  8.76&  8.60&  8.43&&$16.8\pm1.7$\\
39&J05100427-2340407&&  77.5178& -23.6780&  41.4$\pm$ 2.3& -13.3$\pm$ 1.1&Z12&M3+M3.5&$11.21$&  9.24&  8.58&  8.36&  8.21&  8.06&&$24.3\pm0.3$\\
40&J05142878-1514546&&  78.6199& -15.2485&  34.2$\pm$ 3.3& -13.1$\pm$ 3.4&Z12&M3.5&$13.14$& 10.95& 10.40& 10.10&  9.98&  9.82&&$21.4\pm0.3$\\
41&J05241317-2104427&&  81.0549& -21.0786&  33.3$\pm$ 2.5& -17.1$\pm$ 2.2&Z12&M4&$12.40$& 10.21&  9.60&  9.32&  9.23&  9.05&&$24.5\pm0.3$\\
42&J05241914-1601153&&  81.0798& -16.0209&  16.0$\pm$ 2.5& -34.8$\pm$ 3.5&Z12&M4.5+M5.0&$11.17$&  8.67&  8.13&  7.81&  7.62&  7.42&&$17.5\pm0.6$\\
43&J05254166-0909123&NLTT 15049&  81.4236&  -9.1534&  39.2$\pm$ 8.0&-188.4$\pm$ 8.0&Z12&M3.8+M5$^{\rm{dd}}$&$10.58$&  8.45&  7.88&  7.62&  7.45&  7.30&$20.7\pm2.2^{\rm{dd}}$&$26.3\pm0.3$\\
44&J05332558-5117131&&  83.3566& -51.2870&  43.8$\pm$ 2.1&  25.1$\pm$ 2.1&Z12&K7&$10.62$&  8.99&  8.36&  8.16&  8.06&  8.06&&$19.6\pm0.4$\\
45&J05335981-0221325&&  83.4992&  -2.3590&  12.3$\pm$ 1.2& -61.3$\pm$ 2.4&Z12&M3&$10.57$&  8.56&  7.88&  7.70&  7.53&  7.43&&$20.9\pm0.2$\\
46&J05392505-4245211&&  84.8544& -42.7559&  40.8$\pm$ 1.3&  17.5$\pm$ 1.9&Z12&M2&$11.34$&  9.45&  8.80&  8.60&  8.47&  8.38&&$21.9\pm0.2$\\
47&J05395494-1307598&&  84.9789& -13.1333&  20.3$\pm$ 4.8& -11.7$\pm$ 5.4&Z12&M3&$12.62$& 10.60&  9.98&  9.72&  9.61&  9.48&&$24.9\pm0.4$\\
48&J05470650-3210413&&  86.7771& -32.1782&  23.7$\pm$ 0.9&   7.1$\pm$ 1.7&Z12&M2.5&$11.91$&  9.86&  9.22&  9.03&  8.92&  8.79&&$21.9\pm0.6$\\
49&J05575096-1359503&&  89.4624& -13.9973&   0.0$\pm$ 5.0&   0.0$\pm$ 5.0&F16&M7$^{\rm{dd}}$&& 12.87& 12.15& 11.73& 11.24& 10.60&&$30.3\pm2.8^{\rm{dd}}$\\
50&J06045215-3433360&AP Col&  91.2173& -34.5600&  27.3$\pm$ 0.3& 340.9$\pm$ 0.3&Ri11&M5$^{\rm{r}}$&$9.60^{\rm{x}}$&  7.74&  7.18&  6.87&  6.67&  6.39&$8.4\pm0.1^{\rm{x}}$&$22.4\pm0.3^{\rm{x}}$\\
51&J06085283-2753583&&  92.2201& -27.8995&   8.9$\pm$ 3.5&  10.7$\pm$ 3.5&F16&M8.5e$^{\rm{r}}$&& 13.60& 12.90& 12.37& 11.98& 11.62&$31.3\pm3.5^{\rm{j}}$&$24.0\pm1.0^{\rm{w}}$\\
52&J06112997-7213388&&  92.8749& -72.2274&  23.2$\pm$ 1.6&  60.2$\pm$ 1.7&Z12&M4+M5&$11.83$&  9.55&  8.96&  8.70&  8.55&  8.36&&$18.2\pm2.0$\\
53&J06131330-2742054&&  93.3055& -27.7015& -13.1$\pm$ 1.6&  -0.3$\pm$ 1.3&Z12&M3.5&$10.17$&  8.00&  7.43&  7.14&  7.01&  6.82&$29.4\pm0.9^{\rm{y}}$&$22.5\pm0.2$\\
54&J06434532-6424396&& 100.9388& -64.4110&   1.6$\pm$ 2.4&  53.1$\pm$ 2.4&Z12&M3+M4+M5&$11.31$&  9.29&  8.59&  8.37&  8.24&  8.09&&$20.2\pm0.4$\\
55&J08173943-8243298&& 124.4143& -82.7249& -80.3$\pm$ 1.1& 102.5$\pm$ 0.8&Z12&M3.5+&$9.08^{\rm{i}}$&  7.47&  6.84&  6.59&  6.48&  6.27&&$15.6\pm1.5$\\
56&J08471906-5717547&& 131.8294& -57.2985&-123.0$\pm$ 1.2&  12.3$\pm$ 1.2&Z12&M4&$11.57$&  9.41&  8.81&  8.55&  8.37&  8.19&&$30.2\pm0.2$\\
57&J10260210-4105537&& 156.5088& -41.0983& -45.3$\pm$ 1.4&  -2.5$\pm$ 1.4&Z12&M0.5&$11.09$&  9.18&  8.49&  8.27&  8.15&  8.06&&\\
58&J10285555+0050275&HIP 51317& 157.2315&   0.8410&-603.8$\pm$ 1.9&-728.9$\pm$ 2.0&vL07&M2V$^{\rm{r}}$&$7.39^{\rm{a}}$&  6.18&  5.61&  5.31&  5.18&  4.87&$7.07\pm0.03^{\rm{ff}}$&$8.3\pm0.5^{\rm{m}}$\\
59&J11115267-4401538&& 167.9695& -44.0316& -22.0$\pm$ 2.0& -12.0$\pm$ 4.0&Z05&M3.9$^{\rm{cc}}$&$13.65^{\rm{i}}$& 12.09& 11.49& 11.22& 11.10& 10.91&&$17.6\pm0.3^{\rm{cc}}$\\
60&J11305355-4628251&& 172.7231& -46.4737& -35.3$\pm$ 2.2&   4.7$\pm$ 1.8&Z12&M2.4$^{\rm{cc}}$&$14.13$& 12.09& 11.57& 11.29& 11.14& 10.99&&$10.0\pm0.1^{\rm{cc}}$\\
61&J11592786-4510192&& 179.8661& -45.1720& -52.8$\pm$ 5.1& -12.8$\pm$ 2.8&Z12&M4.5$^{\rm{z}}$&$11.53^{\rm{i}}$&  9.93&  9.35&  9.06&  8.92&  8.72&&\\
62&J12210499-7116493&& 185.2708& -71.2804& -42.7$\pm$ 1.8& -10.2$\pm$ 1.6&Z12&K7&$10.57$&  9.09&  8.42&  8.24&  8.17&  8.17&&$13.5\pm0.3$\\
63&J12265135-3316124&TWA 32& 186.7140& -33.2701& -54.0$\pm$ 5.9& -35.0$\pm$ 6.3&2MAW&M6.3$^{\rm{cc}}$&$13.44$& 10.69& 10.12&  9.78&  9.57&  9.21&$15.1\pm0.7^{\rm{h}}$&\\
64&J12300521-4402359&& 187.5217& -44.0433& -56.8$\pm$ 7.0& -12.8$\pm$ 1.9&Z12&M4$^{\rm{z}}$&$12.65$& 10.45&  9.84&  9.57&  9.44&  9.26&&\\
65&J12383713-2703348&& 189.6547& -27.0597&-185.1$\pm$ 5.1&-185.2$\pm$ 5.1&Ro10&M2.5+&$10.57$&  8.73&  8.08&  7.84&  7.66&  7.57&&$9.9\pm0.2$\\
66&J14284804-7430205&& 217.2002& -74.5057& -61.6$\pm$ 1.7& -34.6$\pm$ 1.7&Z12&M1$^{\rm{v,d}}$&$11.07$&  9.26&  8.57&  8.35&  8.26&  8.21&&$11.0\pm0.6$\\
67&J14361471-7654534&& 219.0613& -76.9149& -45.0$\pm$ 1.9& -17.4$\pm$ 1.9&Z12&M0.5&$11.69$&  9.84&  9.17&  8.96&  8.83&  8.75&&\\
68&J15244849-4929473&& 231.2021& -49.4965&-120.8$\pm$ 8.0&-241.0$\pm$ 8.0&Z12&M2&$9.45^{\rm{i}}$&  8.16&  7.53&  7.30&  7.14&  7.02&&$10.3\pm0.2$\\
69&J15310958-3504571&& 232.7899& -35.0825& -20.6$\pm$ 2.0& -25.4$\pm$ 2.0&Z12&M4.5$^{\rm{z}}$&$12.40^{\rm{i}}$& 10.72& 10.10&  9.80&  9.63&  9.40&&\\
70&J16430128-1754274&& 250.7554& -17.9076& -26.6$\pm$ 1.2& -52.4$\pm$ 1.3&Z12&M0.5&$11.18$&  9.44&  8.76&  8.55&  8.44&  8.41&&$-9.3\pm0.4$\\
\enddata
\end{deluxetable}
\floattable
\begin{deluxetable}{hchcccccccccccccc}%[htbp]
\renewcommand\thetable{1}
\tabletypesize{\footnotesize}
\rotate
\tablecaption{\textit{continued}}
\label{tab:stardata}
\tablehead{
\nocolhead{Index}&\multicolumn{2}{c}{2MASS Designation}&\multicolumn{2}{c}{Coordinates}&\multicolumn{3}{c}{Proper Motion}&\colhead{Sp.Type\tablenotemark{a}$^{,}$\tablenotemark{b}}&\multicolumn{6}{c}{Magnitudes}&\colhead{Trigonometric}&\colhead{Radial Velocity\tablenotemark{b}}\\
\cline{6-8}
\cline{10-15}
\nocolhead{}&\colhead{}&\nocolhead{Other}&\colhead{$\alpha$}&\colhead{$\delta$}&\colhead{$\mu_{\alpha}\cos{\delta}$}&\colhead{$\mu_{\delta}$}&\colhead{Ref.\tablenotemark{b}}&\colhead{(Opt.)}&\colhead{$I$\tablenotemark{b}}&\colhead{$J$}&\colhead{$H$}&\colhead{$K_S$}&\colhead{$W1$}&\colhead{$W2$}&\colhead{distance\tablenotemark{b}}&\colhead{}\\
\cline{11-13}
\nocolhead{}&\colhead{}&\nocolhead{}&\colhead{(J2000.0)}&\colhead{(J2000.0)}&\colhead{(mas yr$^{-1}$)}&\colhead{(mas yr$^{-1}$)}&\colhead{}&\colhead{}&\colhead{}&\multicolumn{3}{c}{(2MASS)}&\colhead{}&\colhead{}&\colhead{(pc)}&\colhead{(km s$^{-1}$)}
}
\startdata
71&J16572029-5343316&& 254.3346& -53.7255& -13.0$\pm$ 6.3& -85.1$\pm$ 2.2&Z12&M3&$10.61$&  8.69&  8.07&  7.79&  7.68&  7.57&&$1.4\pm0.2$\\
72&J18420694-5554254&& 280.5290& -55.9071&   9.7$\pm$12.1& -81.2$\pm$ 2.8&Z12&M3.5&$11.61$&  9.49&  8.82&  8.58&  8.49&  8.33&&$0.3\pm0.5$\\
73&J19225071-6310581&& 290.7113& -63.1828&  -7.9$\pm$16.7& -77.5$\pm$ 1.9&Z12&M3&$11.41$&  9.45&  8.82&  8.58&  8.43&  8.29&&$6.4\pm1.5$\\
74&J19355595-2846343&& 293.9832& -28.7762&  34.0$\pm$12.0& -58.0$\pm$12.0&F16&M9 $\gamma$$^{\rm{k}}$&& 13.95& 13.18& 12.71& 12.38& 11.90&&\\
75&J19560294-3207186&& 299.0123& -32.1219&  35.2$\pm$ 1.8& -59.9$\pm$ 1.5&Z12&M4+&$11.03$&  8.96&  8.34&  8.11&  7.92&  7.76&&$-3.7\pm2.2$\\
76&J20004841-7523070&2MUCD 20845& 300.2018& -75.3853&  69.0$\pm$12.0&-110.0$\pm$ 4.0&F16&M9$^{\rm{bb}}$&& 12.73& 11.97& 11.51& 11.13& 10.81&&$4.4\pm2.8^{\rm{k}}$\\
77&J20013718-3313139&& 300.4049& -33.2206&  27.0$\pm$ 3.2& -58.6$\pm$ 2.0&Z12&M1&$10.85$&  9.15&  8.46&  8.24&  8.16&  8.09&&$-3.7\pm0.2$\\
78&J20100002-2801410&& 302.5001& -28.0281&  40.7$\pm$ 3.0& -62.0$\pm$ 1.7&Z12&M2.5+M3.5&$10.92$&  8.65&  8.01&  7.73&  7.61&  7.45&$48.0\pm3.1^{\rm{y}}$&$-5.8\pm0.6$\\
79&J20333759-2556521&& 308.4066& -25.9478&  52.8$\pm$ 1.7& -75.9$\pm$ 1.3&Z12&M4.5&$12.42$&  9.71&  9.15&  8.88&  8.68&  8.44&$48.3\pm3.3^{\rm{y}}$&$-7.6\pm0.4$\\
80&J20465795-0259320&& 311.7415&  -2.9922&  53.0$\pm$ 2.5&-109.5$\pm$ 1.7&Z12&M0&$10.75$&  9.12&  8.44&  8.27&  8.24&  8.22&&$-14.2\pm0.3$\\
81&J21100535-1919573&& 317.5223& -19.3326&  89.0$\pm$ 0.9& -89.9$\pm$ 1.8&Z12&M2&$10.07$&  8.11&  7.45&  7.20&  7.02&  7.00&&$-5.7\pm0.4$\\
82&J21265040-8140293&& 321.7100& -81.6748&  55.6$\pm$ 1.4&-101.8$\pm$ 3.0&F16&L3 $\gamma$$^{\rm{g}}$&& 15.54& 14.40& 13.55& 12.93& 12.47&$32.0\pm2.7^{\rm{k}}$&$10.0\pm0.5^{\rm{k}}$\\
83&J21471964-4803166&& 326.8318& -48.0546&  50.9$\pm$ 1.7& -74.0$\pm$ 2.0&Z12&M4&$13.08$& 10.73& 10.19&  9.92&  9.75&  9.60&&$10.4\pm2.9$\\
84&J21521039+0537356&HIP 107948& 328.0433&   5.6266& 128.1$\pm$ 7.0&-135.6$\pm$26.7&2MAW&M2Ve*$^{\rm{r}}$&$9.75^{\rm{gg}}$&  8.25&  7.65&  7.39&  7.14&  7.07&$30.5\pm5.3^{\rm{ff}}$&$-15.1\pm1.5^{\rm{dd}}$\\
85&J22021626-4210329&& 330.5677& -42.1758&  50.4$\pm$ 1.0& -90.9$\pm$ 1.5&Z12&M1&$10.72$&  8.93&  8.23&  7.99&  7.89&  7.87&&$-2.6\pm0.5$\\
86&J22440873-5413183&& 341.0364& -54.2218&  70.7$\pm$ 1.3& -60.0$\pm$ 1.3&Z12&M4+&$11.51$&  9.36&  8.71&  8.47&  8.30&  8.14&&$1.6\pm1.6$\\
87&J22470872-6920447&& 341.7863& -69.3458&  70.9$\pm$ 1.6& -58.9$\pm$ 1.8&Z12&K7(sb1)$^{\rm{v,s}}$&$10.37$&  8.89&  8.30&  8.09&  8.01&  8.00&&$17.3\pm0.2$\\
88&J23131671-4933154&& 348.3196& -49.5543&  77.5$\pm$ 2.1& -88.1$\pm$ 1.7&Z12&M4&$12.07$&  9.76&  9.14&  8.92&  8.77&  8.58&&$1.9\pm0.3$\\
89&J23221088-0301417&& 350.5453&  -3.0283&  92.4$\pm$ 1.6& -68.3$\pm$ 1.7&Z12&K7&$10.44$&  8.73&  8.12&  7.93&  7.85&  7.89&&$-5.4\pm0.3$\\
90&J23285763-6802338&& 352.2402& -68.0427&  66.8$\pm$ 1.9& -67.1$\pm$ 1.7&Z12&M2.5&$11.27$&  9.26&  8.64&  8.38&  8.27&  8.16&&$10.8\pm3.4$\\
91&J23301341-2023271&& 352.5559& -20.3909& 311.8$\pm$ 3.2&-207.4$\pm$ 3.0&vL07&M3(sb2)$^{\rm{v,ee}}$&$9.02^{\rm{ff}}$&  7.20&  6.61&  6.33&  6.23&  6.02&$16.2\pm0.9^{\rm{ff}}$&$-5.7\pm0.8$\\
92&J23320018-3917368&& 353.0008& -39.2936& 193.4$\pm$17.9&-178.4$\pm$17.9&Ro10&M3&$11.08$&  8.90&  8.26&  8.02&  7.88&  7.75&&$11.1\pm0.2$\\
93&J23452225-7126505&& 356.3427& -71.4474&  80.3$\pm$ 2.2& -62.4$\pm$ 2.1&Z12&M3.5&$12.40$& 10.19&  9.57&  9.32&  9.17&  8.98&&$8.6\pm0.3$\\
94&J23474694-6517249&& 356.9456& -65.2903&  79.2$\pm$ 1.2& -66.8$\pm$ 1.2&Z12&M1.5&$10.88$&  9.10&  8.39&  8.17&  8.09&  8.02&&$6.2\pm0.5$\\
\enddata
\tablenotetext{\rm{a}}{The symbols $\beta$ and $\gamma$ are used when referring to \citet{Allers2013}  INT-G and VL-G gravity classes, for simplicity. }
\tablenotetext{\rm{b}}{References~: Spectral type references from \citet{Riaz2006} unless otherwise stated. $I$ magnitude from \citet{Zacharias2012} unless otherwise stated. Radial velocities from \citet{Malo2014a} unless otherwise stated. (a)~\citealt{Anderson2012},~(b)~\citealt{Bailey2012},~(c)~\citealt{Bergfors2010},~(d)~\citealt{Bowler2015},~(e)~\citealt{Cruz2003},~(f)~\citealt{Cruz2007},~(g)~\citealt{Cruz2009},~(h)~Donaldson et al., in prep.,~(i)~\citealt{Epchtein1997},~(j)~\citealt{Faherty2012},~(k)~\citealt{Faherty2016},~(l)~Gagn\'{e} (2017, private communication),~(m)~\citealt{Gontcharov2006},~(n)~\citealt{Janson2012},~(o)~\citealt{Kirkpatrick2010},~(p)~\citealt{Kiss2011},~(q)~\citealt{Koen2010},~(r)~\citealt{Malo2013},~(s)~\citealt{Malo2014a},~(t)~\citealt{Reid2002},~(u)~\citealt{Reid2008b},~(v)~\citealt{Riaz2006},~(w)~\citealt{Rice2010},~(x)~\citealt{Riedel2011},~(y)~\citealt{Riedel2014},~(z)~\citealt{Rodriguez2011},~(aa)~\citealt{Roeser2010},~(bb)~\citealt{Schmidt2007},~(cc)~\citealt{Shkolnik2011},~(dd)~\citealt{Shkolnik2012},~(ee)~\citealt{Torres2006},~(ff)~\citealt{vanLeeuwen2007},~(gg)~\citealt{Weis1991},~(hh)~\citealt{Zacharias2005},~(ii)~\citealt{Zacharias2012},~(F16)~\citealt{Faherty2016},~(Ri11)~\citealt{Riedel2011},~(Ro10)~\citealt{Roeser2010},~(Z05)~\citealt{Zacharias2005},~(Z12)~\citealt{Zacharias2012},~(vL07)~\citealt{vanLeeuwen2007},~(2MAW)~Measured from 2MASS and WISE. }
\end{deluxetable}
%\end{landscape}
%Dans cette version de la table, j'ai enleve la mention /OLD à J0440, ajouté le \citealt{..} pour Allers2013 dans sa reference, enlevé le ``?'' du HLC de J0349, enlevé le ``?'' du BF de J1226 (source Weinberger:2012gk, Donaldson in prep), enlevé le ``?'' du AY de J2147
%For High Likelihood Candidates
%tout ajuster pour ajouter une colonne vide entre adopted age range et adopted distance range (pour que les clines ne se touchent pas
%Enlevé la mention ; NC: no constrain on age. dans la légende et changé l'étoile qui était identifiée comme tel pour NYI
% Remplacé Donaldson et al. in prep par  \citealt{Donaldson:2016uk}
%\rm{a} pour les tablenote en bas de la table (pour pas que ce soit en italique)
\floattable
\begin{deluxetable*}{hccccccccc}%[htbp]
\renewcommand\thetable{2}
\rotate
\tablecaption{Sample Age and Distance}
\label{tab-staragedist}
\tablehead{
\nocolhead{Index}&\colhead{2MASS designation}&\colhead{Status\tablenotemark{a}}&\multicolumn{3}{c}{Adopted Age Range\tablenotemark{b}}&&\multicolumn{3}{c}{Adopted Distance Range\tablenotemark{c}}\\
\nocolhead{}&\colhead{}&\colhead{}&\multicolumn{3}{c}{(Myr)}&&\multicolumn{3}{c}{(pc)}\\
\cline{4-6}
\cline{8-10}
\nocolhead{Index}&\colhead{}&\colhead{}&\colhead{min}&\colhead{max}&\colhead{constraints}&&\colhead{min}&\colhead{max}&\colhead{source}
}
\startdata
0&J00040288-6410358&HLC&    41&    49&THA&&    43&    49&\ds; THA\\
1&J00172353-6645124&HLC&    21&    27&BPMG&&    36&    41&$\pi$; \citealt{Riedel2014}\\
2&J00325584-4405058&AY&    41&   200&THA; ABDMG&&    30&    61&$\pi$; \citealt{Faherty2016}\\
3&J00374306-5846229&YO&     5&   200&YO&&    38&    60&\dsp\\
4&J01071194-1935359&YO&    21&   200&YO; Li&&    13&    69&\ds; BPMG; COL; FIELD\\
5&J01123504+1703557&HLC&   130&   200&ABDMG&&    45&    49&\ds; ABDMG\\
6&J01132958-0738088&YO&     5&  1000&YO; H$\alpha$&&    39&    59&\ds; FIELD\\
7&J01220441-3337036&HLC&    41&    49&THA&&    37&    41&\ds; THA\\
8&J01351393-0712517&AY&    21&    48&COL; BPMG&&    35&    40&$\pi$; \citealt{Shkolnik2012}\\
9&J01415823-4633574&HLC&    41&    49&THA&&    37&    42&\ds; THA\\
10&J01484087-4830519&HLC&   130&   200&ABDMG&&    34&    38&\ds; ABDMG\\
11&J01521830-5950168&HLC&    41&    49&THA&&    37&    41&\ds; THA\\
12&J02045317-5346162&HLC&    41&    49&THA&&    39&    43&\ds; THA\\
13&J02070176-4406380&HLC&    41&    49&THA&&    41&    45&\ds; THA\\
14&J02155892-0929121&HLC&    41&    49&THA&&    41&    45&\ds; THA\\
15&J02215494-5412054&HLC&    41&    49&THA&&    36&    41&\ds; THA\\
16&J02224418-6022476&HLC&    41&    49&THA&&    29&    33&\ds; THA\\
17&J02251947-5837295&C&    41&    49&THA&&    40&    45&\ds; THA\\
18&J02303239-4342232&HLC&    38&    48&COL&&    50&    54&\ds; COL\\
19&J02340093-6442068&HLC&    41&    49&THA&&    42&    49&\ds; THA\\
20&J02485260-3404246&AY&    38&    49&COL; THA&&    40&    46&\ds; COL; THA\\
21&J02564708-6343027&AY&    38&    49&COL; THA&&    50&    60&\ds; COL; THA\\
22&J03050976-3725058&HLC&    38&    48&COL&&    68&    76&\ds; COL\\
23&J03350208+2342356&BF&    21&    27&BPMG&&    40&    44&$\pi$; \citealt{Shkolnik2012}\\
24&J03494535-6730350&HLC&    38&    48&COL&&    77&    85&\ds; COL\\
25&J04082685-7844471&HLC&    38&    56&CAR&&    53&    55&\ds; CAR\\
26&J04091413-4008019&HLC&    38&    48&COL&&    58&    68&\ds; COL\\
27&J04213904-7233562&HLC&    41&    49&THA&&    49&    57&\ds; THA\\
28&J04240094-5512223&HLC&    38&    48&COL&&    62&    72&\ds; COL\\
29&J04363294-7851021&HLC&   130&   200&ABDMG&&    51&    61&\ds; ABDMG\\
30&J04365738-1613065&AY&    21&    49&THA; BPMG&&    12&    34&\ds; THA; BPMG\\
31&J04402325-0530082&NYI&   200& 10000&\citealt{Allers2013}; \citealt{Cruz2009}&&     9&     9&$\pi$; \citealt{Riedel2014}\\
32&J04433761+0002051&HLC&    21&    27&BPMG&&    22&    28&\ds; BPMG\\
33&J04440099-6624036&HLC&    41&    49&THA&&    50&    58&\ds; THA\\
34&J04480066-5041255&HLC&    41&    49&THA&&    48&    56&\ds; THA\\
\enddata
\end{deluxetable*}

\floattable
\begin{deluxetable*}{hccccccccc}%[htbp]
\renewcommand\thetable{2}
\rotate
\tablecaption{\textit{continued}}
\tablehead{
\nocolhead{Index}&\colhead{2MASS designation}&\colhead{Status\tablenotemark{a}}&\multicolumn{3}{c}{Adopted Age Range\tablenotemark{b}}&&\multicolumn{3}{c}{Adopted Distance Range\tablenotemark{c}}\\
\nocolhead{}&\colhead{}&\colhead{}&\multicolumn{3}{c}{(Myr)}&&\multicolumn{3}{c}{(pc)}\\
\cline{4-6}
\cline{8-10}
\nocolhead{Index}&\colhead{}&\colhead{}&\colhead{min}&\colhead{max}&\colhead{constraints}&&\colhead{min}&\colhead{max}&\colhead{source}
}
\startdata
35&J04533054-5551318&BF&   130&   200&ABDMG&&    10&    11&$\pi$; \citealt{vanLeeuwen2007}\\
36&J04571728-0621564&HLC&   130&   200&ABDMG&&    42&    48&\ds; ABDMG\\
37&J04593483+0147007&BF&    21&    27&BPMG&&    24&    27&$\pi$; \citealt{vanLeeuwen2007}\\
38&J05090356-4209199&AY&    21&    50&BPMG; ARG&&    19&    55&\ds; BPMG; ARG\\
39&J05100427-2340407&HLC&    38&    48&COL&&    44&    54&\ds; COL\\
40&J05142878-1514546&HLC&    38&    48&COL&&    54&    66&\ds; COL\\
41&J05241317-2104427&HLC&    38&    48&COL&&    46&    56&\ds; COL\\
42&J05241914-1601153&HLC&    21&    27&BPMG&&    14&    24&\ds; BPMG\\
43&J05254166-0909123&HLC&   130&   200&ABDMG&&    18&    22&$\pi$; \citealt{Shkolnik2012}\\
44&J05332558-5117131&HLC&    41&    49&THA&&    48&    56&\ds; THA\\
45&J05335981-0221325&HLC&    21&    27&BPMG&&    30&    38&\ds; BPMG\\
46&J05392505-4245211&AY&    38&    49&COL; THA&&    37&    56&\ds; COL; THA\\
47&J05395494-1307598&HLC&    38&    48&COL&&    59&    77&\ds; COL\\
48&J05470650-3210413&HLC&    38&    48&COL&&    45&    59&\ds; COL\\
49&J05575096-1359503&YO&     5&   400&YO&&    30&    49&\dph; \citealt{Shkolnik2012}\\
50&J06045215-3433360&BF&    30&    50&ARG&&     8&     8&$\pi$; \citealt{Riedel11}\\
51&J06085283-2753583&YO&     5&   200&YO&&    20&    32&\dsp\\
52&J06112997-7213388&HLC&    38&    56&CAR&&    45&    49&\ds; CAR\\
53&J06131330-2742054&HLC&    21&    27&BPMG&&    28&    30&$\pi$; \citealt{Riedel2014}\\
54&J06434532-6424396&AY&    38&    56&CAR; COL&&    49&    59&\ds; CAR; COL\\
55&J08173943-8243298&HLC&    21&    27&BPMG&&    25&    29&\ds; BPMG\\
56&J08471906-5717547&HLC&   130&   200&ABDMG&&    20&    24&\ds; ABDMG\\
57&J10260210-4105537&C&     7&    13&TWA&&    56&    66&\ds; TWA\\
58&J10285555+0050275&BF&   130&   200&ABDMG&&     7&     7&$\pi$; \citealt{vanLeeuwen2007}\\
59&J11115267-4401538&YO&    90&   160&\citealt{Shkolnik2011}&&    27&    40&\dph; \citealt{Shkolnik2011}\\
60&J11305355-4628251&YO&    20&   130&\citealt{Shkolnik2011}&&    49&    74&\dph; \citealt{Shkolnik2011}\\
61&J11592786-4510192&YO&     5&    12&ScoCen; \citealt{Rodriguez2011}&&    44&    66&\dph; \citealt{Rodriguez2011}\\
62&J12210499-7116493&YO&     3&    15&\citealt{Kiss2011}&&    88&   107&dkin; \citealt{Kiss2011}\\
63&J12265135-3316124&BF&     7&    13&TWA&&    63&    69&$\pi$; \citealt{Donaldson:2016uk}\\
64&J12300521-4402359&YO&     5&    12&ScoCen; \citealt{Rodriguez2011}&&    55&    82&\dph; \citealt{Rodriguez2011}\\
65&J12383713-2703348&HLC&   130&   200&ABDMG&&    22&    24&\ds; ABDMG\\
66&J14284804-7430205&YO&    21&  1000&No Li; Malo (2017, in preparation); H$\alpha$; \citealt{Riaz2006}&&    24&    68&\ds; BPMG; CAR; FIELD\\
67&J14361471-7654534&YO&    21&  1000&No Li; Malo (2017, in preparation); H$\alpha$; \citealt{Riaz2006}&&    26&    44&\ds; FIELD\\
68&J15244849-4929473&HLC&   130&   200&ABDMG&&    23&    25&\ds; ABDMG\\
69&J15310958-3504571&YO&     5&    12&ScoCen; \citealt{Rodriguez2011}& &   56&    84&\dph; \citealt{Rodriguez2011}\\
70&J16430128-1754274&YO&    21&   200&Li; Malo (2017, in preparation)&&    31&    51&\ds; FIELD\\
71&J16572029-5343316&HLC&    21&    27&BPMG&&    49&    55&\ds; BPMG\\
\enddata
\end{deluxetable*}

\floattable
\begin{deluxetable*}{hccccccccc}%[htbp]
\renewcommand\thetable{2}
\rotate
\tablecaption{\textit{continued}}
\tablehead{
\nocolhead{Index}&\colhead{2MASS designation}&\colhead{Status\tablenotemark{a}}&\multicolumn{3}{c}{Adopted Age Range\tablenotemark{b}}&&\multicolumn{3}{c}{Adopted Distance Range\tablenotemark{c}}\\
\nocolhead{}&\colhead{}&\colhead{}&\multicolumn{3}{c}{(Myr)}&&\multicolumn{3}{c}{(pc)}\\
\cline{4-6}
\cline{8-10}
\nocolhead{Index}&\colhead{}&\colhead{}&\colhead{min}&\colhead{max}&\colhead{constraints}&&\colhead{min}&\colhead{max}&\colhead{source}
}
\startdata
72&J18420694-5554254&HLC&    21&    27&BPMG&&    49&    57&\ds; BPMG\\
73&J19225071-6310581&AY&    21&    49&BPMG; THA&&    49&    66&\ds; BPMG; THA\\
74&J19355595-2846343&YO&     5&   200&YO&&    24&    38&\dsp\\
75&J19560294-3207186&HLC&    21&    27&BPMG&&    54&    62&\ds; BPMG\\
76&J20004841-7523070&HLC&    21&    27&BPMG&&    28&    35&\ds; BPMG\\
77&J20013718-3313139&HLC&    21&    27&BPMG&&    58&    66&\ds; BPMG\\
78&J20100002-2801410&HLC&    21&    27&BPMG&&    44&    51&$\pi$; \citealt{Riedel2014}\\
79&J20333759-2556521&HLC&    21&    27&BPMG&&    44&    51&$\pi$; \citealt{Riedel2014}\\
80&J20465795-0259320&HLC&   130&   200&ABDMG&&    44&    48&\ds; ABDMG\\
81&J21100535-1919573&HLC&    21&    27&BPMG&&    31&    35&\ds; BPMG\\
82&J21265040-8140293&YO&     5&   200&YO&&    29&    34&$\pi$; \citealt{Faherty2016}\\
83&J21471964-4803166&AY&    21&   200&ABDMG; BPMG; THA&&    41&    69&\ds; ABDMG; BPMG; THA\\
84&J21521039+0537356&BF&   130&   200&ABDMG&&    25&    35&$\pi$; \citealt{vanLeeuwen2007}\\
85&J22021626-4210329&HLC&    41&    49&THA&&    43&    49&\ds; THA\\
86&J22440873-5413183&HLC&    41&    49&THA&&    45&    51&\ds; THA\\
87&J22470872-6920447&HLC&   130&   200&ABDMG&&    52&    58&\ds; ABDMG\\
88&J23131671-4933154&HLC&    41&    49&THA&&    38&    42&\ds; THA\\
89&J23221088-0301417&YO&    10&  1000&YO; H$\alpha$&&    30&    46&\ds; COL; BPMG\\
90&J23285763-6802338&HLC&    41&    49&THA&&    45&    51&\ds; THA\\
91&J23301341-2023271&HLC&    38&    48&COL&&    15&    17&$\pi$; \citealt{vanLeeuwen2007}\\
92&J23320018-3917368&HLC&   130&   200&ABDMG&&    22&    24&\ds; ABDMG\\
93&J23452225-7126505&HLC&    41&    49&THA&&    42&    48&\ds; THA\\
94&J23474694-6517249&HLC&    41&    49&THA&&    44&    48&\ds; THA\\
\enddata
\tablenotetext{\rm{a}}{Status: BF: bona fide; HLC: high-likelihood candidate, unambiguous membership (high probability considering radial velocity or parallax measurements; C: Candidate (high probability without RV or plx confirmation); AY: ambiguous young (more than one association has a high probability); YO: other young stars; NYI: no youth indicator.}
\tablenotetext{\rm{b}}{For high-likelihood candidates and stars with ambiguous membership, the total range of the association(s) is given.}
\tablenotetext{\rm{c}}{Adopted distance range source: \ds: statistical distance; \dph: photometric distance; \dsp: spectrophotometric distance; dkin: kinematic distance; $\pi$: parallax.}
\end{deluxetable*}

CAR: $45^{+11}_{-7}$\,Myr. For ARG, \citet{Bell2015} did not assign a final age, arguing that the list of members appears to be contaminated. According to their analysis, it is unclear that the members represent a single coeval population. Assessing whether this association is indeed a unique ensemble of coeval objects is beyond the scope of this paper, so the age range determined by \citet{Makarov2000} (30--50\,Myr) is used for ARG objects.

For YO stars, other age indicators  were used to constrain the age of the star. Several low-mass stars from the \citet{Riaz2006} sample and analyzed by \BANYAN\ for moving group membership have $H_{\alpha}$ emission measurements. Since $H_{\alpha}$ in emission remains for \env1\,Gyr for early M dwarfs \citep{West2008}, this sets an upper age limit for these stars. The presence of lithium was also used to constrain the age of some stars. For some stars analyzed by BANYAN II (M7 or later types), the gravity classes of \citet{Allers2013} were used. \citet{Allers2013} have constructed a gravity classification scheme based on several spectral indices in the near-infrared that allows us to classify low-mass stars and brown dwarfs in one of three categories: field gravity (FLD-G), intermediate gravity (INT-G), and very low gravity (VL-G). The INT-G and VL-G gravity classes were built to correspond, respectively, to the $\beta$ and $\gamma$ visual classifications introduced by \citet{Cruz2009} and used in the spectral types listed in Table~1. The three classes respectively correspond to objects of decreasing surface gravities and thus likely decreasing ages. Using a sample of age-calibrated objects, they determined that the VL-G class corresponds to an age range of \env10--30\,Myr, and that the INT-G class corresponds to an age range of \env50--200\,Myr. They note that there are exceptions, but there is an observed trend where the fraction of VL-G objects with respect to INT-G or FLD-G objects is higher in younger moving groups \citep{Allers2013, Faherty2016}. When no other age constraints were available, spectral indices were used to assess if they belong to one of the two low-gravity classes. If it was the case, the stars were assigned 200\,Myr as an upper bound; if not, they were assigned 200\,Myr as a lower bound. When a lower or upper bound was not available for age, the values 5\,Myr and 10,000\,Myr were respectively conservatively assigned, assuming the stars are not in star-forming regions and do not belong to the thick disk or halo. Table~2 summarizes the adopted age range for all survey targets. The midrange age was computed for each star. The median of the midrange ages is \env45\,Myr.

\subsubsection{Distance}
Trigonometric distances are used when available. This is the case for all BF stars, by definition. For HLC stars that do not have a trigonometric distance measurement, the statistical distance in the most probable association is used. For AY stars, the total range of statistical distances in the associations that have high membership probabilities is assigned. 
For YO stars that do not benefit from a parallax measurement, the spectrophotometric distance (\dsp) was estimated from the method of \citet{Gagne2015b}. Spectral types listed in Table~1 were used in combination with the spectral-type absolute-magnitude sequences of \env5--200\,Myr objects in a specific near-infrared (NIR) band to obtain a distance estimate and measurement error for a given object. These measurements were performed on the 2MASS $J$, $H$ and $K_S$ bands and the AllWISE $W1$ and $W2$ bands and were each represented by a Gaussian probability density function (PDF) with the appropriate central position and characteristic width. The five PDFs were then multiplied together to obtain a final measurement PDF; the maximum position of this PDF corresponds to the most probable distance, and the 68\% range corresponds to measurement uncertainties. This method does not account for correlations between the different NIR magnitudes of young objects and may thus slightly underestimate the measurement errors (see \citealp{Gagne2015b} for more detail). Table~2 and Figure~\ref{fig:star-prop-histo_2} summarize the adopted distance ranges. The median distance of the sample is \env45\,pc. 
%For the other YO stars, a photometric distance (\dph) was often already available in the literature. 

%%%%%%%%%%%%%%%%%%%%%%%%%%%%%%%%%%%%%%%%%%%%%%%%%%%%%%%%%%%%%%%%%%%%%%%%

\section{Observation and Data Reduction}
\label{obs}
%%%%%%%%%%%%%%%%%%%%%%%%%%%%%%%%%%
\subsection{Observing Strategy}
\label{strat}
\begin{figure*}[htbp]
\begin{center}
\includegraphics[width=15cm]{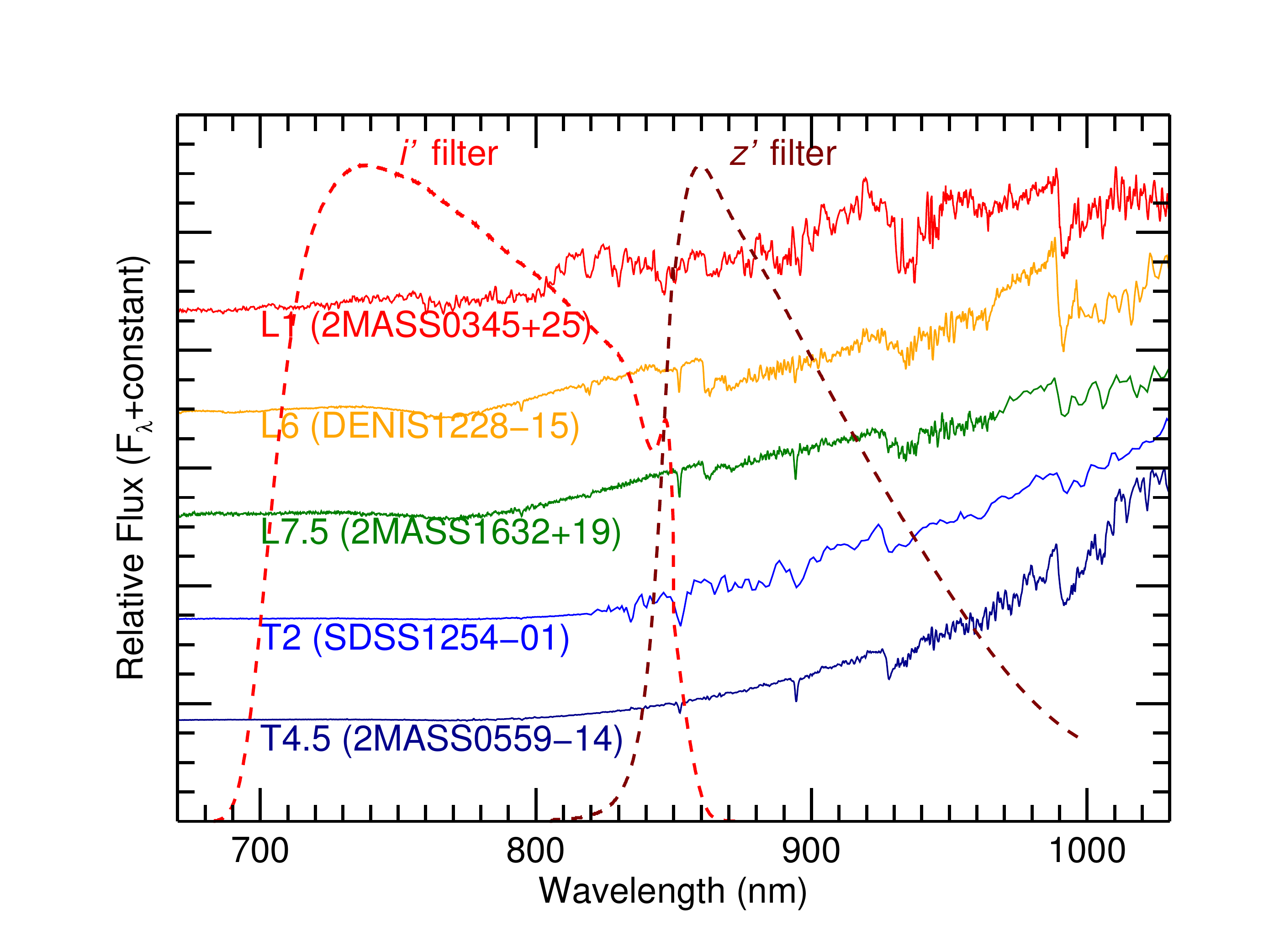}
%\includegraphics[width=8.5cm]{f2b.eps}
%izfilterspectra.eps}
%Sptype_iz_Zhang2009.eps}
\caption{The far-red spectra of five objects with spectral types ranging from early-Ls to mid-Ts (from the L and T dwarf data archive; \url{http://staff.gemini.edu/~sleggett/LTdata.html}). The spectra are normalized at 960\,nm and offset for clarity. The transmission curves of the GMOS \i\ and \z\ filters (similar to SDSS filters) are superimposed. The \z\ filter curve includes the response from the detector. }
\label{fig:iz}
\end{center}
\end{figure*}

In this survey, planetary-mass companions are identified via their distinctively high \iz\ color. This strategy was previously used to identify a number of T dwarfs in the Canada-France Brown Dwarf Survey \citep{Delorme2008b,Albert2011}. This is because low-mass objects give off most of their flux in the infrared. Figure~\ref{fig:iz} shows that the rise of the flux around 780\,nm in the SED of brown dwarfs is steeper for late spectral types, which results in an \iz\ increasing from \iz\env2 for types earlier than L4 to \iz\env3 for L8 and \iz\env4 for T3 (\citealp{Zhang2009}). %Polynomial fit of the \iz\ color (SDSS) for ultra-cool dwarfs presented by \citet{Zhang2009}. There is a rapid increase of the \iz\ color in the L0--T5 range that is useful to distinguish these objects from most other astrophysical objects.

\begin{figure*}[htbp]
\begin{center}
\includegraphics[width=15cm]{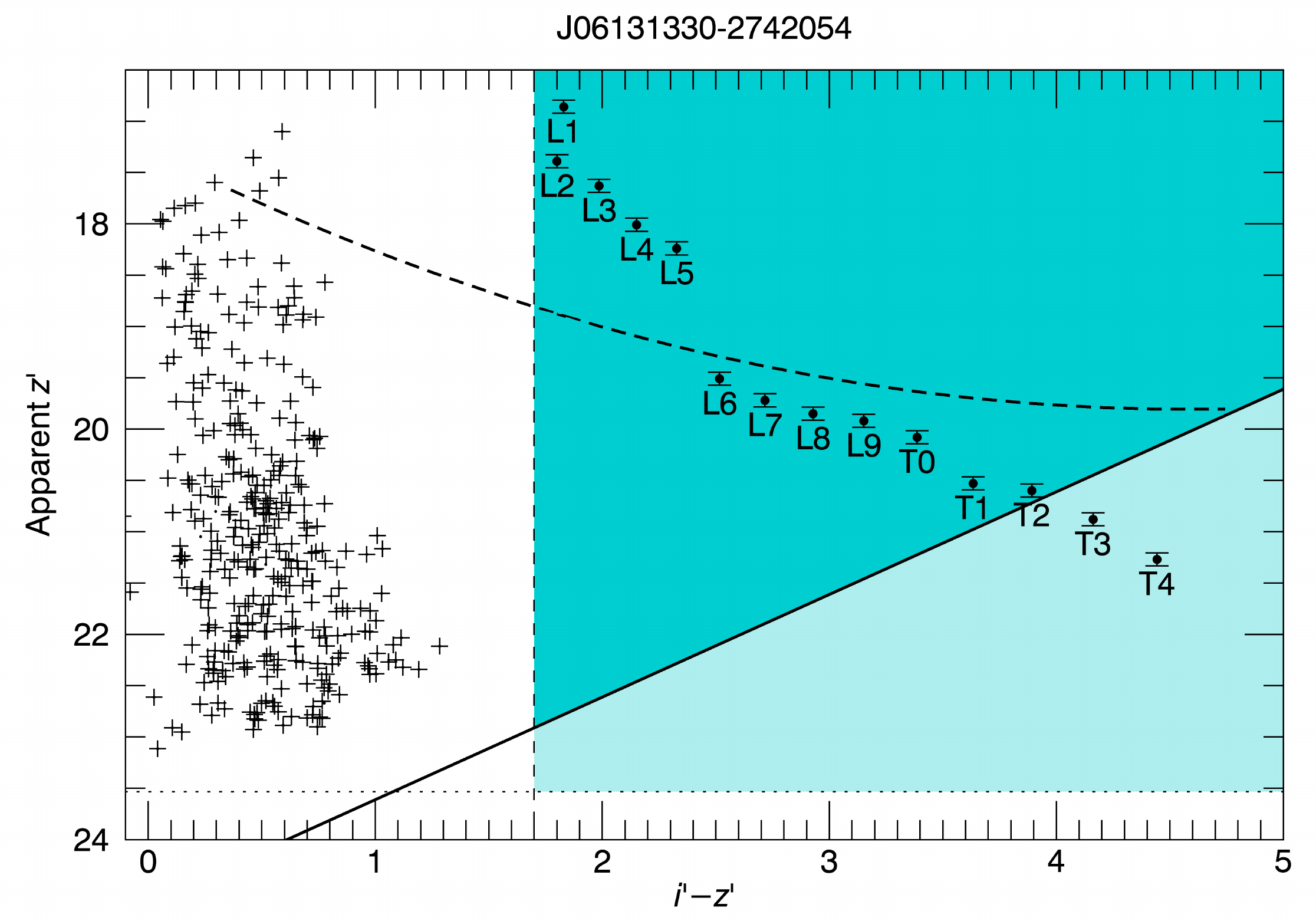}%f3.pdf}
%zvsiz-2png.pdf} 
\caption{Color-magnitude diagram for all objects present in the field of a typical target of the survey. Also shown are fields L0--T4 at the range of distance of the target \citep{West2005,Zhang2009}. Younger objects with inflated radii would appear higher (brighter) on this figure. There are 353 objects identified in this field, but none with an \iz\gta1.3. Objects in the dark cyan region are detected in both the \i\ and \z-band while cooler objects, down to T4, are detected as \i dropouts (light cyan region). The dashed line indicates the approximate limit above which objects are also detected in 2MASS ($J<16$, earlier than L5).} 
\label{fig:zvsiz}
\end{center}
\end{figure*}

Figure~\ref{fig:zvsiz} shows the apparent \z\ magnitude versus \iz\ for all objects identified in the field of one of the targets, 2MASS~J06131330-2742054. Typical fields L0--T4 are also shown, with the apparent magnitudes they would have at the mean distance of the target, 29\,pc \citep{West2005,Zhang2009}. For each spectral type, the dot corresponds to the value of a field object. Younger objects are expected to have inflated radii \citep{Chabrier2000} and would thus appear slightly brighter and thus higher on the figure. The vast majority of objects in a given field are much bluer (to the left) than the \iz=1.7 threshold adopted. Very few false positives are thus expected. Besides young low-mass companions, the only objects that have such colors are field L/T dwarfs and the much rarer high-redshift quasars \citep{Delorme2008b,Reyle2010}. Field L/T dwarfs are rare. \citet{Allen2005} have estimated a local density of L dwarfs ($M_J$=11.75--14.75) to be $7.35\times10^{-3}$pc$^{-3}$, while \citet{Reyle2010} estimated the local density of T0--T5.5 dwarfs to be $1.4\times10^{-3}$pc$^{-3}$. Within a 5\farcm5 FOV and a maximum distance of 100 pc, each field samples \env0.85\,pc$^{3}$. For the entire survey (81\,pc$^{3}$), that amounts to \env0.6 L dwarfs and \env0.11 early-T. Less than one such false positive was therefore expected. An astrometric follow-up can be made to confirm common proper motion to the primary and eliminate these false positives. Host stars in the present sample are nearby and in general have high proper motions. Common proper motion can be detected within at most a few years for all targets. The dashed line in Figure~\ref{fig:zvsiz} indicates the approximate limit above which objects are also detected in the 2MASS catalog \citep{Cutri2003}, calculated using typical \z$-J$ colors \citep{Zhang2009} and the $J<16$ limit of 2MASS. The earliest candidates can thus be readily identified as comoving with the primary, because 2MASS observations were taken \env10 years earlier. High-redshift quasars are even rarer per unit surface at a given apparent magnitude and can be distinguished with broadband NIR photometry. Their flux is not rising toward the infrared (their red color in \iz\ is due to the Lyman forest absorption blueward of the Ly$\alpha$ emission line), and they have much more neutral \z$-J$ colors than substellar companions and would not display common proper motion with the nearby star. Optical and mid-infrared (\textit{WISE}) colors can also help to distinguish those. 

Candidates warmer than \env T2 are detected in both the \i\ and \z-bands, while cooler objects down to \env T4 are detected as \i dropouts (dark and light cyan regions, respectively, on Figure~\ref{fig:iz}). Note that the \i\ and \z\ observations are optimal to identify late-L to early-T companions, which at the young age of the stars in the survey are planetary-mass or low-mass brown dwarfs. Contrary to what would be the case for standard high-contrast imaging surveys, this survey is much \textit{less} sensitive to earlier-L or late-M, which have less distinctive \iz. The focus here is thus on planetary-mass companions and not on brown dwarfs.

The observations allow us to detect companions as close as 5\arcsec --70\arcsec\ to the target (depending on its brightness) and up to the edge of the GMOS 5$\farcm$5 field of view (\env165\arcsec\ from the target). For a typical target at 45\,pc, this allows to survey a distance of \env7400\,au. We chose to limit our analysis to 5000\,au to be complete for most of the targets of the survey and because it corresponds to the observed upper limit on the separation of low-mass stellar binaries \citep{Artigau2007, Caballero2007, Radigan2009, Dhital2010}.

\subsection{Observations}
\label{Obs}
The observations were carried out in 2011--2012 at Gemini South during three different semesters (see Table~3). Broadband imaging was performed with GMOS in the \i\ (iG0327, 700-850\,nm) and \z\ (zG0328, $>$850\,nm) filters. The GMOS detector is made of three 2048$\times$4608 CCDs, with a pixel scale of 0\farcs073/pixel, for a total field of view of 5$\farcm$5 squared. In each band, at least three exposures were taken, with a small dither between each, in order to remove cosmic rays and fill the gaps between the detectors. The exposure time in \z\ (200\,s per individual exposure) was chosen to reach $z=22$, the apparent magnitude of an $M_{z}=18$ object for the most distant targets in the sample (\env80\,pc). This allows us to detect objects down to a temperature of about 900\,K (T5). In the \i\ band, individual exposures of 300\,s were obtained in order to reach \i=24.5 and thus minimally detect objects with \iz=2.5 (\env L6). This constraint on \iz\ minimizes the number of false positives and thus the follow-up time. Observations in the \i\ and \z\ bands were scheduled together when possible, in order to lower the overall time required per observation and reduce the likelihood of astrophysical false positives from variable objects. Observations in both filters typically required $\sim$36 minutes per target, including overheads. A summary of observations for individual targets is shown in Table~4.

\begin{table}[htbp]
\renewcommand\thetable{3}
\caption{Observing Log}
\label{table:obs-campain}
\begin{center}
\begin{tabular}{ccccc}
\hline \hline   
\rule{0pt}{3ex} 
Program no. &  Dates & Total  &  Targets \\
 &   &Time (hr)&  observed \\
\hline
GS-2011B-Q-74&2011 Aug.--Oct.2&22&34\\%2011-08-04 to 2011-10-18
GS-2012A-Q-78&2012 Feb.--Jul.&22.2&27\\%2012-02-03 to 2012-07-29
GS-2012B-Q-75&2012 Jul.--2013 Jan.&20.9&34\\%2012-07-25 to 2013-01-06
\hline
\end{tabular}
\end{center}
\end{table}

%%%%%%%%%%%%%%%%%%%%%%%%%%%%%%%%%%
\subsection{Data Reduction}
\label{data-reduction}

A custom data-reduction pipeline was used to process GMOS \i\ and \z\ images. Each \i\ or \z\ image is composed of three files that correspond to the three 2048$\times$4608 chips of the GMOS detector. After making a basic reduction, including the identification of bad pixels and saturated pixels, overscan and bias subtraction, fringe correction and flat-field division, the astrometry of each portion was independently anchored to the USNO-B1 catalog. The positions of the left and right chips relative to the middle one were then computed for all images using reference points. The median relative position was adopted, and the final \i\ or \z\ images were reconstructed. 

For each star and each filter, three or more images were taken. As optimal photometric conditions were not requested for the observations, the transmission sometimes varied significantly during exposures. The maximal cloud cover requested (CC=70\%) implies patchy clouds or extended thin cirrus clouds that lead to a maximum loss of 0.3 mag\footnote{\url{http://www.gemini.edu/sciops/telescopes-and-sites/observing-condition-constraints}}. Images with a transmission below 70\% of the best case were rejected. If there were more than three images satisfying this condition, all images with a measured FWHM no larger than 1.2 times that of the third best were kept (to avoid adding images with a good transmission but taken under bad seeing). For all stars and in both filters, there were always at least two images remaining. All images were scaled to mach the zero point of the highest-throughput image before median-combining them to obtain a deep image for each filter. 
Table~4 lists, for each object, the number of images that were considered and the FWHM of the combined image produced. The FWHM varies between 0\farcs5 and 1\farcs6 in both filters, with a median of 1\farcs0.
%Observations were mostly made in Band 3, which means that the conditions of observations were not always photometric.

%%%%%%%%%%%%%%%%%%%%%%%%%%%%%%%%%%
\subsection{Assessment of conditions and photometric calibration}
\label{photcal}
One significant challenge in analyzing nonphotometric observations is to flux-calibrate the data. It is useful first to identify which observations were likely taken under photometric conditions and which were not. This can be done by looking at the variation of the transmission in the three \i\ or \z\ images. If the rms of the transmission of consecutive retained images was more than 3\%, the conditions were suspected to be nonphotometric. The fields with nonphotometric conditions were identified with the mention ``light clouds'' (lc) in Table~4. The other were assumed to have been taken under almost photometric conditions (phot). It is possible, although unlikely, that a nonnegligible cloud cover remained stable over a \env20 minutes of observation. That would lead to a slight underestimation of the error on the zero points in those cases. The effect on the results of the survey is however negligible.

When available, the zero point was determined through a cross-match with the Sloan Digital Sky Survey (SDSS DR9; \citealp{Ahn2012}). Other fields were flux-calibrated using the SkyMapper \citep{Wolf2016} early data release\footnote{\url{http://skymapper.anu.edu.au/data-release/}} or the Pan-STARRS \citep{Schlafly2012,Magnier2013} PV3 release. SkyMapper and Pan-STARRS magnitudes were first converted to SDSS magnitudes using, respectively, the procedure explained on the web site\footnote{\url{http://skymapper.anu.edu.au/filter-transformations/}} and the color correction from \citet{Tonry2012}. For each field, point sources are then identified in the calibrated survey field and in that of GMOS. The zero point adopted for each field and filter is the median of the zero points computed for each source, which is the difference between the cataloged magnitude and that computed in the GMOS field. The errors for the zero points computed this way are taken to be the standard deviation of the zero points computed for every source divided by the square root of the number of sources (typically $<0.05$). The medians of zero points obtained from the three surveys are in agreement. The computed zero points for the different fields vary between 26.5 and 27.1 with a median of 26.8 in \i\, and between 25.2 and 26 with a median of 25.7 in \z\ (see Table~4). 

About one-half of the 95 fields are not found in SDSS, Pan-STARRS or SkyMapper and cannot be directly calibrated. For these, the median of the values found for the calibrated fields was assigned. The calibrated fields that were identified non photometric were not used in the computation of this median. An error of 0.15 or 0.25 was conservatively assigned on the zero point assigned this way for observations taken under photometric conditions and nonphotometric conditions, respectively, given the dispersion of the zero points for the fields that were calibrated. This is consistent with the computed $\Delta(\rm{ZP}_{\rm{computed}}-\rm{ZP}_{\rm{median}})$ for the fields for which the zero point was computed and is also compatible with the expected maximal loss of flux under a CC of 70\%. 

%Changé le texte des tablenote a la mitaine!
% Enlevé le - de Sky Mapper
%ajouté renewcommand\thetable{4}
%\rm{a} pour les tablenote en bas de la table (pour pas que ce soit en italique)
%Reformulation de la note ``b'':``Source of the zero point, fields calibrated with SDSS, Sky Mapper and Pan-STARRS are identified SDSS, SM, PS, respectively. Those without a direct calibration are identified as med, since the median of the zero points for all calibrated fields with photometric observations was assigned in those cases.''
%remplacé \asec par \arcsec
\begin{deluxetable*}{hcccccccc}%{hcc>{\centering}p{2.5cm}cccc}
\tablecaption{Summary of individual target observations}
\renewcommand\thetable{4}
\label{tab-obs}
\tablehead{
\nocolhead{Index} & \colhead{Name} & \colhead{Filter} & \colhead{Obs. Date(s; UT)} & \colhead{$N_{\rm{exp}}$} & \colhead{Condition\tablenotemark{a}} & \colhead{FWHM}  & \colhead{Zero point} & \colhead{Source\tablenotemark{b}} \\
\nocolhead{} & \colhead{} & \colhead{} & \colhead{(YYYYMMDD)} & \colhead{}  & \colhead{}   & \colhead{\arcsec}  & \colhead{} & \colhead{}}
\startdata
 0&J00040288-6410358&$i$&20120920&3&phot&0.9&26.87$\pm$ 0.15&med\\
&&$z$&20120920&3&lc&0.8&25.75$\pm$ 0.25&med\\
 1&J00172353-6645124&$i$&20110804&3&phot&1.2&26.87$\pm$ 0.15&med\\
&&$z$&20110804&3&phot&1.1&25.75$\pm$ 0.15&med\\
 2&J00325584-4405058&$i$&20120921&3&phot&1.4&26.87$\pm$ 0.15&med\\
&&$z$&20120921&4&phot&1.3&25.75$\pm$ 0.15&med\\
 3&J00374306-5846229&$i$&20120920&3&phot&1.1&26.87$\pm$ 0.15&med\\
&&$z$&20120920&3&lc&1.1&25.75$\pm$ 0.25&med\\
 4&J01071194-1935359&$i$&20111006&3&phot&1.4&27.00$\pm$ 0.07&PS\\
&&$z$&20111006&3&lc&1.1&26.03$\pm$ 0.07&PS\\
 5&J01123504+1703557&$i$&20110922,20111018&7&phot&1.1&26.87$\pm$ 0.01&SDSS\\
&&$z$&20110922,20111018&3&phot&1.0&25.73$\pm$ 0.02&SDSS\\
 6&J01132958-0738088&$i$&20111007&3&phot&1.4&27.04$\pm$ 0.03&SDSS\\
&&$z$&20111007&3&phot&1.3&25.86$\pm$ 0.02&SDSS\\
 7&J01220441-3337036&$i$&20111005&3&phot&1.6&26.87$\pm$ 0.15&med\\
&&$z$&20111005&3&phot&1.5&25.75$\pm$ 0.15&med\\
 8&J01351393-0712517&$i$&20110922&3&phot&1.2&27.01$\pm$ 0.02&SDSS\\
&&$z$&20110922&3&phot&1.1&25.88$\pm$ 0.04&SDSS\\
 9&J01415823-4633574&$i$&20120915&3&phot&1.3&26.87$\pm$ 0.15&med\\
&&$z$&20120915&3&lc&1.2&25.75$\pm$ 0.25&med\\
10&J01484087-4830519&$i$&20111006&3&phot&1.2&26.91$\pm$ 0.04&SM\\
&&$z$&20111006&3&phot&1.1&25.79$\pm$ 0.05&SM\\
11&J01521830-5950168&$i$&20120203&3&phot&1.1&26.87$\pm$ 0.15&med\\
&&$z$&20120203&3&phot&1.1&25.75$\pm$ 0.15&med\\
12&J02045317-5346162&$i$&20110804&3&phot&1.2&26.87$\pm$ 0.15&med\\
&&$z$&20110804&3&phot&1.1&25.75$\pm$ 0.15&med\\
13&J02070176-4406380&$i$&20110921&3&lc&0.9&26.51$\pm$ 0.05&SM\\
&&$z$&20110921&3&lc&0.8&25.64$\pm$ 0.03&SM\\
14&J02155892-0929121&$i$&20111007&3&phot&1.6&26.91$\pm$ 0.02&SDSS\\
&&$z$&20111007&3&phot&1.4&25.73$\pm$ 0.04&SDSS\\
15&J02215494-5412054&$i$&20120827&3&lc&0.9&26.87$\pm$ 0.25&med\\
&&$z$&20120827&3&lc&0.9&25.75$\pm$ 0.25&med\\
16&J02224418-6022476&$i$&20110916&3&phot&1.3&26.87$\pm$ 0.15&med\\
&&$z$&20110916&3&phot&1.2&25.75$\pm$ 0.15&med\\
17&J02251947-5837295&$i$&20120826&3&phot&1.6&26.87$\pm$ 0.15&med\\
&&$z$&20120826&3&phot&1.6&25.75$\pm$ 0.15&med\\
18&J02303239-4342232&$i$&20120730&3&phot&1.1&26.87$\pm$ 0.15&med\\
&&$z$&20120730&3&phot&0.8&25.75$\pm$ 0.15&med\\
19&J02340093-6442068&$i$&20120826&3&phot&1.2&26.78$\pm$ 0.04&SM\\
&&$z$&20120826&3&lc&1.3&25.71$\pm$ 0.03&SM\\
20&J02485260-3404246&$i$&20111006&3&phot&1.1&26.87$\pm$ 0.15&med\\
&&$z$&20111006&3&phot&1.0&25.75$\pm$ 0.15&med\\
21&J02564708-6343027&$i$&20111005&2&phot&1.5&26.87$\pm$ 0.15&med\\
&&$z$&20111005&3&phot&1.3&25.75$\pm$ 0.15&med\\
22&J03050976-3725058&$i$&20111007,20111011&6&phot&1.5&26.97$\pm$ 0.04&SM\\
&&$z$&20111007,20111011&3&phot&1.6&25.83$\pm$ 0.04&SM\\
23&J03350208+2342356&$i$&20130103&3&phot&1.1&26.84$\pm$ 0.02&PS\\
&&$z$&20130103&3&phot&1.0&25.70$\pm$ 0.02&PS\\
24&J03494535-6730350&$i$&20111009&3&phot&0.9&26.87$\pm$ 0.15&med\\
&&$z$&20111009&3&phot&0.8&25.75$\pm$ 0.15&med\\
25&J04082685-7844471&$i$&20111011&3&phot&1.0&26.87$\pm$ 0.15&med\\
&&$z$&20111011&3&phot&1.1&25.75$\pm$ 0.15&med\\
26&J04091413-4008019&$i$&20120827&3&phot&1.3&26.87$\pm$ 0.15&med\\
&&$z$&20120827&3&phot&1.3&25.75$\pm$ 0.15&med\\
27&J04213904-7233562&$i$&20110922&3&phot&1.2&26.87$\pm$ 0.15&med\\
&&$z$&20110922&3&phot&1.1&25.75$\pm$ 0.15&med\\
28&J04240094-5512223&$i$&20110916&3&phot&1.3&26.87$\pm$ 0.15&med\\
&&$z$&20110916&3&phot&1.0&25.75$\pm$ 0.15&med\\
29&J04363294-7851021&$i$&20120915&3&lc&1.3&26.53$\pm$ 0.02&SM\\
&&$z$&20120915&3&lc&1.2&25.20$\pm$ 0.02&SM\\
30&J04365738-1613065&$i$&20111008&3&phot&1.4&26.98$\pm$ 0.02&PS\\
&&$z$&20111008&3&phot&1.4&25.78$\pm$ 0.01&PS\\
31&J04402325-0530082&$i$&20120921&3&phot&0.8&27.01$\pm$ 0.02&SDSS\\
&&$z$&20120921&3&phot&0.9&25.83$\pm$ 0.02&SDSS\\
32&J04433761+0002051&$i$&20121010&3&phot&0.8&26.85$\pm$ 0.02&SDSS\\
&&$z$&20121010&3&phot&0.9&25.70$\pm$ 0.02&SDSS\\
33&J04440099-6624036&$i$&20110918&3&phot&0.9&26.87$\pm$ 0.15&med\\
&&$z$&20110918&3&phot&0.9&25.75$\pm$ 0.15&med\\
34&J04480066-5041255&$i$&20121019&3&phot&1.4&26.87$\pm$ 0.15&med\\
&&$z$&20121019&3&phot&1.3&25.75$\pm$ 0.15&med\\
35&J04533054-5551318&$i$&20120921&3&phot&1.1&26.87$\pm$ 0.15&med\\
&&$z$&20120921&3&phot&1.5&25.75$\pm$ 0.15&med\\
36&J04571728-0621564&$i$&20111018&3&phot&1.0&26.87$\pm$ 0.03&PS\\
&&$z$&20111018&3&phot&0.9&25.73$\pm$ 0.02&PS\\
37&J04593483+0147007&$i$&20121025&3&phot&1.4&26.83$\pm$ 0.03&PS\\
&&$z$&20121025&3&phot&1.4&25.75$\pm$ 0.02&PS\\
38&J05090356-4209199&$i$&20120826,20121011&3&phot&1.3&26.82$\pm$ 0.05&SM\\
&&$z$&20120826,20121011&3&phot&1.2&25.69$\pm$ 0.02&SM\\
39&J05100427-2340407&$i$&20121019&3&phot&1.4&26.88$\pm$ 0.03&SM\\
&&$z$&20121019&3&phot&1.3&25.79$\pm$ 0.01&SM\\
40&J05142878-1514546&$i$&20111009&3&phot&0.8&26.90$\pm$ 0.02&PS\\
&&$z$&20111009&3&phot&0.7&25.81$\pm$ 0.03&PS\\
41&J05241317-2104427&$i$&20111009&3&phot&0.7&26.93$\pm$ 0.02&SM\\
&&$z$&20111009&3&phot&0.6&25.78$\pm$ 0.02&SM\\
42&J05241914-1601153&$i$&20121025&3&phot&1.3&26.82$\pm$ 0.02&PS\\
&&$z$&20121025&3&phot&1.3&25.75$\pm$ 0.02&PS\\
43&J05254166-0909123&$i$&20121221&3&phot&0.9&26.83$\pm$ 0.03&PS\\
&&$z$&20121221&3&phot&0.9&25.72$\pm$ 0.01&PS\\
44&J05332558-5117131&$i$&20110922&3&phot&1.1&26.87$\pm$ 0.15&med\\
&&$z$&20110922&3&phot&1.2&25.75$\pm$ 0.15&med\\
45&J05335981-0221325&$i$&20121025&3&phot&1.2&26.80$\pm$ 0.01&SDSS\\
&&$z$&20121025&3&phot&1.3&25.71$\pm$ 0.01&SDSS\\
46&J05392505-4245211&$i$&20111006&3&phot&0.9&26.87$\pm$ 0.15&med\\
&&$z$&20111006&3&phot&0.9&25.75$\pm$ 0.15&med\\
47&J05395494-1307598&$i$&20121016&3&phot&0.9&26.89$\pm$ 0.02&PS\\
&&$z$&20121016&3&phot&0.9&25.83$\pm$ 0.01&PS\\
48&J05470650-3210413&$i$&20121012&3&phot&1.1&26.87$\pm$ 0.15&med\\
&&$z$&20121012&3&phot&1.0&25.75$\pm$ 0.15&med\\
49&J05575096-1359503&$i$&20121010&3&phot&0.9&26.90$\pm$ 0.02&PS\\
&&$z$&20121010&3&phot&0.9&25.83$\pm$ 0.02&PS\\
50&J06045215-3433360&$i$&20121023&4&phot&1.3&26.87$\pm$ 0.15&med\\
&&$z$&20121023&3&phot&1.1&25.75$\pm$ 0.15&med\\
51&J06085283-2753583&$i$&20121012&3&phot&1.0&26.87$\pm$ 0.02&SM\\
&&$z$&20121012&3&phot&0.9&25.77$\pm$ 0.01&SM\\
52&J06112997-7213388&$i$&20121221&3&phot&1.1&26.75$\pm$ 0.02&SM\\
&&$z$&20121221&3&phot&1.0&25.64$\pm$ 0.02&SM\\
53&J06131330-2742054&$i$&20111009&3&phot&0.8&26.98$\pm$ 0.02&SM\\
&&$z$&20111009&3&phot&0.8&25.84$\pm$ 0.01&SM\\
54&J06434532-6424396&$i$&20110923&3&phot&0.8&26.84$\pm$ 0.04&SM\\
&&$z$&20110923&3&phot&0.8&25.74$\pm$ 0.02&SM\\
55&J08173943-8243298&$i$&20130103&3&phot&1.0&26.87$\pm$ 0.15&med\\
&&$z$&20130103&3&phot&1.0&25.75$\pm$ 0.15&med\\
56&J08471906-5717547&$i$&20130103&3&phot&1.0&26.66$\pm$ 0.01&SM\\
&&$z$&20130103&3&phot&1.0&25.61$\pm$ 0.01&SM\\
57&J10260210-4105537&$i$&20120203&3&phot&0.8&26.76$\pm$ 0.02&SM\\
&&$z$&20120203&3&phot&0.8&25.75$\pm$ 0.01&SM\\
58&J10285555+0050275&$i$&20120204&3&phot&1.2&27.07$\pm$ 0.03&SDSS\\
&&$z$&20120204&3&phot&0.9&25.90$\pm$ 0.03&SDSS\\
59&J11115267-4401538&$i$&20120203&3&phot&0.7&26.87$\pm$ 0.15&med\\
&&$z$&20120203&3&phot&0.8&25.75$\pm$ 0.15&med\\
60&J11305355-4628251&$i$&20120205&3&phot&0.6&26.87$\pm$ 0.15&med\\
&&$z$&20120205&3&phot&0.6&25.75$\pm$ 0.15&med\\
61&J11592786-4510192&$i$&20120204&3&phot&0.6&26.68$\pm$ 0.03&SM\\
&&$z$&20120204&3&phot&0.5&25.64$\pm$ 0.02&SM\\
62&J12210499-7116493&$i$&20120205&3&phot&0.5&26.87$\pm$ 0.15&med\\
&&$z$&20120205&3&phot&0.5&25.75$\pm$ 0.15&med\\
63&J12265135-3316124&$i$&20120204&3&phot&0.7&26.87$\pm$ 0.15&med\\
&&$z$&20120204&3&phot&0.7&25.75$\pm$ 0.15&med\\
64&J12300521-4402359&$i$&20120204&3&phot&0.6&26.87$\pm$ 0.15&med\\
&&$z$&20120204&3&phot&0.6&25.75$\pm$ 0.15&med\\
65&J12383713-2703348&$i$&20120204&3&phot&0.7&26.85$\pm$ 0.02&SM\\
&&$z$&20120204&3&phot&0.7&25.81$\pm$ 0.01&SM\\
66&J14284804-7430205&$i$&20120205&3&phot&0.6&26.52$\pm$ 0.04&SM\\
&&$z$&20120205&3&phot&0.6&25.67$\pm$ 0.01&SM\\
67&J14361471-7654534&$i$&20120205&3&phot&0.6&26.87$\pm$ 0.15&med\\
&&$z$&20120205&3&phot&0.5&25.75$\pm$ 0.15&med\\
68&J15244849-4929473&$i$&20120229&3&phot&0.7&26.87$\pm$ 0.15&med\\
&&$z$&20120229&3&phot&0.7&25.75$\pm$ 0.15&med\\
69&J15310958-3504571&$i$&20120229&3&phot&0.8&26.87$\pm$ 0.15&med\\
&&$z$&20120229&3&phot&0.8&25.75$\pm$ 0.15&med\\
70&J16430128-1754274&$i$&20120304&3&lc&0.7&26.79$\pm$ 0.01&PS\\
&&$z$&20120304&3&phot&0.7&25.66$\pm$ 0.01&PS\\
71&J16572029-5343316&$i$&20120305&3&phot&0.7&26.87$\pm$ 0.15&med\\
&&$z$&20120305&3&lc&0.7&25.75$\pm$ 0.25&med\\
72&J18420694-5554254&$i$&20120318&3&phot&1.3&26.74$\pm$ 0.01&SM\\
&&$z$&20120318&3&phot&1.3&25.63$\pm$ 0.01&SM\\
73&J19225071-6310581&$i$&20120318&3&phot&1.3&26.69$\pm$ 0.02&SM\\
&&$z$&20120318&3&phot&1.2&25.64$\pm$ 0.02&SM\\
74&J19355595-2846343&$i$&20120726&3&phot&1.0&26.81$\pm$ 0.01&PS\\
&&$z$&20120726&3&phot&1.0&25.71$\pm$ 0.01&PS\\
75&J19560294-3207186&$i$&20120318&3&phot&1.0&26.87$\pm$ 0.15&med\\
&&$z$&20120318&3&phot&1.0&25.75$\pm$ 0.15&med\\
76&J20004841-7523070&$i$&20120726&3&phot&1.4&26.87$\pm$ 0.15&med\\
&&$z$&20120726&3&phot&1.4&25.75$\pm$ 0.15&med\\
77&J20013718-3313139&$i$&20110804&3&lc&0.7&26.87$\pm$ 0.25&med\\
&&$z$&20110804&3&phot&0.8&25.75$\pm$ 0.15&med\\
78&J20100002-2801410&$i$&20110804&3&lc&0.8&26.84$\pm$ 0.01&PS\\
&&$z$&20110804&3&lc&0.7&25.59$\pm$ 0.01&PS\\
79&J20333759-2556521&$i$&20110804&3&lc&0.9&26.78$\pm$ 0.01&PS\\
&&$z$&20110804&3&lc&0.8&25.55$\pm$ 0.01&PS\\
80&J20465795-0259320&$i$&20110804&3&lc&1.1&26.90$\pm$ 0.01&SDSS\\
&&$z$&20110804&2&lc&0.9&25.70$\pm$ 0.01&SDSS\\
81&J21100535-1919573&$i$&20110804&3&phot&0.9&26.97$\pm$ 0.03&PS\\
&&$z$&20110804&3&phot&0.9&25.84$\pm$ 0.02&PS\\
82&J21265040-8140293&$i$&20120726,20120729&4&phot&1.4&26.87$\pm$ 0.15&med\\
&&$z$&20120726,20120729&3&phot&1.4&25.75$\pm$ 0.15&med\\
83&J21471964-4803166&$i$&20110804&3&phot&1.3&26.89$\pm$ 0.04&SM\\
&&$z$&20110804&3&phot&1.1&25.81$\pm$ 0.03&SM\\
84&J21521039+0537356&$i$&20120529&2&lc&1.0&26.56$\pm$ 0.01&SDSS\\
&&$z$&20120529&3&lc&0.9&25.57$\pm$ 0.01&SDSS\\
85&J22021626-4210329&$i$&20110804&3&phot&1.0&26.87$\pm$ 0.15&med\\
&&$z$&20110804&3&phot&1.0&25.75$\pm$ 0.15&med\\
86&J22440873-5413183&$i$&20120514&3&phot&1.1&26.71$\pm$ 0.04&SM\\
&&$z$&20120514&3&phot&1.6&25.71$\pm$ 0.02&SM\\
87&J22470872-6920447&$i$&20120514&3&phot&0.9&26.87$\pm$ 0.15&med\\
&&$z$&20120514&3&phot&1.0&25.75$\pm$ 0.15&med\\
88&J23131671-4933154&$i$&20120514&3&phot&1.4&26.87$\pm$ 0.15&med\\
&&$z$&20120514&3&phot&1.2&25.75$\pm$ 0.15&med\\
89&J23221088-0301417&$i$&20120726&3&phot&1.2&26.79$\pm$ 0.02&SDSS\\
&&$z$&20120726&3&phot&1.2&25.63$\pm$ 0.02&SDSS\\
90&J23285763-6802338&$i$&20120514&3&phot&1.1&26.87$\pm$ 0.15&med\\
&&$z$&20120514&3&phot&1.0&25.75$\pm$ 0.15&med\\
91&J23301341-2023271&$i$&20120726&3&phot&1.6&26.78$\pm$ 0.03&SDSS\\
&&$z$&20120726&3&phot&1.0&25.52$\pm$ 0.05&SDSS\\
92&J23320018-3917368&$i$&20120522&3&phot&1.2&26.88$\pm$ 0.04&SM\\
&&$z$&20120522&3&phot&0.9&25.71$\pm$ 0.03&SM\\
93&J23452225-7126505&$i$&20120522&4&phot&1.0&26.87$\pm$ 0.15&med\\
&&$z$&20120522&4&lc&1.5&25.75$\pm$ 0.25&med\\
94&J23474694-6517249&$i$&20120529&3&phot&1.0&26.87$\pm$ 0.15&med\\
&&$z$&20120529&3&phot&0.8&25.75$\pm$ 0.15&med\\
\enddata
\tablenotetext{\rm{a}}{The observing condition was assigned based on the variation between the three or more exposures in the filter: photometric (phot) if the rms is $<3\%$, or light clouds (lc) otherwise. See text for more details.}
\tablenotetext{\rm{b}}{Source of the zero point fields calibrated with SDSS, SkyMapper, and Pan-STARRS are identified as SDSS, SM, and PS, respectively. Those without a direct calibration are identified as med, since the median of the zero points for all calibrated fields with photometric observations was assigned in those cases.}
\end{deluxetable*}

%%%%%%%%%%%%%%%%%%%%%%%%%%%%%%%%%%%%%%%%%%%%%%%%%%%%%%%%%%%%%%%%%%%%%%%%%

\section{Results}
\label{results}
%%%%%%%%%%%%%%%%%%%%%%%%%%%%%%%%%%

\subsection{Candidate Companions} 
\label{companions}

The flux-calibrated and median-combined \i\ and \z\ images were used to search for companions. All point sources were first identified on the \z\ images using the IDL procedure \texttt{find}. The position of each source was fine-tuned by fitting a 2D gaussian with \texttt{gcntrd}. The same sources were then identified in the \i\ images at the determined sky coordinates using the astrometries of the images. Sources identified in the \z\ image but not in the \i\ image are kept, since late-type candidates are not expected to be found in the \i\ image. The sky-subtracted flux in 1 FWHM apertures (the sky is sampled in an annulus between 2 and 3 FWHM) was determined for all sources in the \i\ and the \z\ images using aperture photometry. This flux was then converted to \i\ and \z\ magnitudes using the zero points determined previously (see Section \ref{photcal}). Sources that are too close to the edges of the images, with an extended PSF, or with saturated flux in \i\ or \z\ images, were excluded. The total number of sources retained varies substantially between the targets, between a few dozen to a few thousand. The \iz\ of the sources was then computed. Only a lower limit for the \iz\ color is available for sources not identified in the \i\ image. 

At 5--10\,Myr, the age of the youngest stars in the sample, the transition between planetary-mass and brown dwarfs takes place around the spectral types L1--L2. According to \citet{West2005}, a typical L1--L2 dwarf has an \iz\ color of about 1.8. Sources with \iz$>$1.7 were thus conservatively selected. As seen in Section \ref{photcal}, there are targets for which the zero points of the \i\ and \z\ images are more uncertain. In the worst cases, the \iz\ is expected to be off by 0.5 mag, considering the errors listed in Table~4. Two approaches were used in order to be sure to identify all plausible planetary-mass companions (with spectral type L0 and later) around these stars. In the first approach, the center of the \iz\ distribution of all sources identified in the field was computed and artificially shifted to 0.5, which is the approximate \iz\ of an early M, the typical star expected in these far-red images \citealp{Hawley2002,West2005}. Then, all sources with (shifted) \iz\ greater than 1.7 were inspected. In the second approach, the peak of the \iz\ was left untouched, but all sources with \iz$>$1.2 were inspected. 

Considering the eccentricity distribution of \citet{Cumming2008} and random viewing time and inclination, it can be shown (see Section \ref{complete}) that $<$2\% of candidates with projected distances $>$8000\,au will have a semimajor axis below 5000\,au. Candidates with projected distances $<$8000\,au from the target were conservatively retained, to make sure all candidates with semimajor axis $<$5000\,au are identified. Most target stars are saturated in the GMOS \z\ image. To find their precise position, the R.A., decl. from the 2MASS catalog listed in Table~1 is used a first approximation. This position does not take into account the proper motion, so it is often off by several pixels (on average six to seven pixels but sometimes as much as several dozens of pixels). For stars that are unsaturated, a 2D Gaussian profile was fitted with IDL function \texttt{mpfit2dfun}; for the other stars, the position was found manually, fitting a circle region on the star.

These selection criteria were found to efficiently reject contaminants and left only a few candidates in any given field. Most of these were easily eliminated by a visual inspection of the median-combined and individual \i\ and \z\ images. Remaining false positives were likely cosmic rays or were located in the diffraction peaks of bright stars that affected their photometry. Some non-point-source objects that were not eliminated automatically with the criteria in the \texttt{find} procedure were also discarded. Other sources fall in part or entirely off the detectors in one or more of the individual images. Finally, the typical L and T colors and magnitudes shown in Figure~\ref{fig:zvsiz} \citep{West2005,Zhang2009} were helpful in discarding objects with \iz$>1.7$ that are much too faint in \z\ to be brown dwarfs at the distance of the source. Only one candidate survived all selection criteria, around the M3 ABDMG star GU Psc (2MASS~J01123504+1703557).

\subsubsection{GU~Psc~b}
GU~Psc has an estimated age range of 130--200\,Myr (given the most recent estimate of ABDMG age from \citealp{Bell2015}), and a corresponding statistical distance range of 45--49\,pc. The characterization of the system is described by \citet{Naud2014} and only summarized here. GU~Psc~b was detected in the \z-band observations of 2011 September 22 ($z_{AB}=21.76\pm0.07$), but not in the \i\ band. Follow-up observations with the same instrument and observational setup were made on 2011 October 18 to obtain a deeper \i-band image; four additional 300 s \i-band images were taken. The new \i-band imaging still did not reveal the companion but provided a 3$\sigma$ upper limit on the flux of \i\ $>$ 25.28, indicating a very red \iz\ color ($>$ 3.5 at $3 \sigma$). The $J_{Vega}=18.15\pm0.05$ was measured at CFHT/WIRCam and the $K_{\rm s}=17.10\pm0.15$ was obtained with the 1.6\,m Telescope of Observatoire du Mont-M\'{e}gantic. A spectrum was obtained with GNIRS at Gemini North, and a spectral type of T3.5 was assigned to the companion. The $J-K_{\rm s}\textsc{(Vega)}=1.05\pm0.16$ is significantly redder than the bulk of field T dwarfs of comparable $z-J$, most likely because of the reduced collision-induced absorption by molecular hydrogen due to a low surface gravity. Using atmosphere models, the temperature and surface gravity were evaluated (\Teff=1000--1100\,K and \logg=4.18--4.36). Using hot-start evolutionary models \citep{Saumon2008, Allard2013}, the mass was estimated to be in the range 9--13\,\MJ. Follow-up $J$-band observations allowed the confirmation of the common proper motion with the primary star, located 42\arcsec (2000\,au) away from it. 
%The original three 300 s exposures in \i and three 200 s exposures in \z were taken on 2011 September 22 (see Figure~\ref{fig:composite}).  

%%%%%%%%%%%%%%%%%%%%%%%%%%%%%%%%%%
\subsection{Detection limits} 
\label{contrast}

\begin{figure*}[htbp]
\begin{center}
\includegraphics[width=8.8cm]{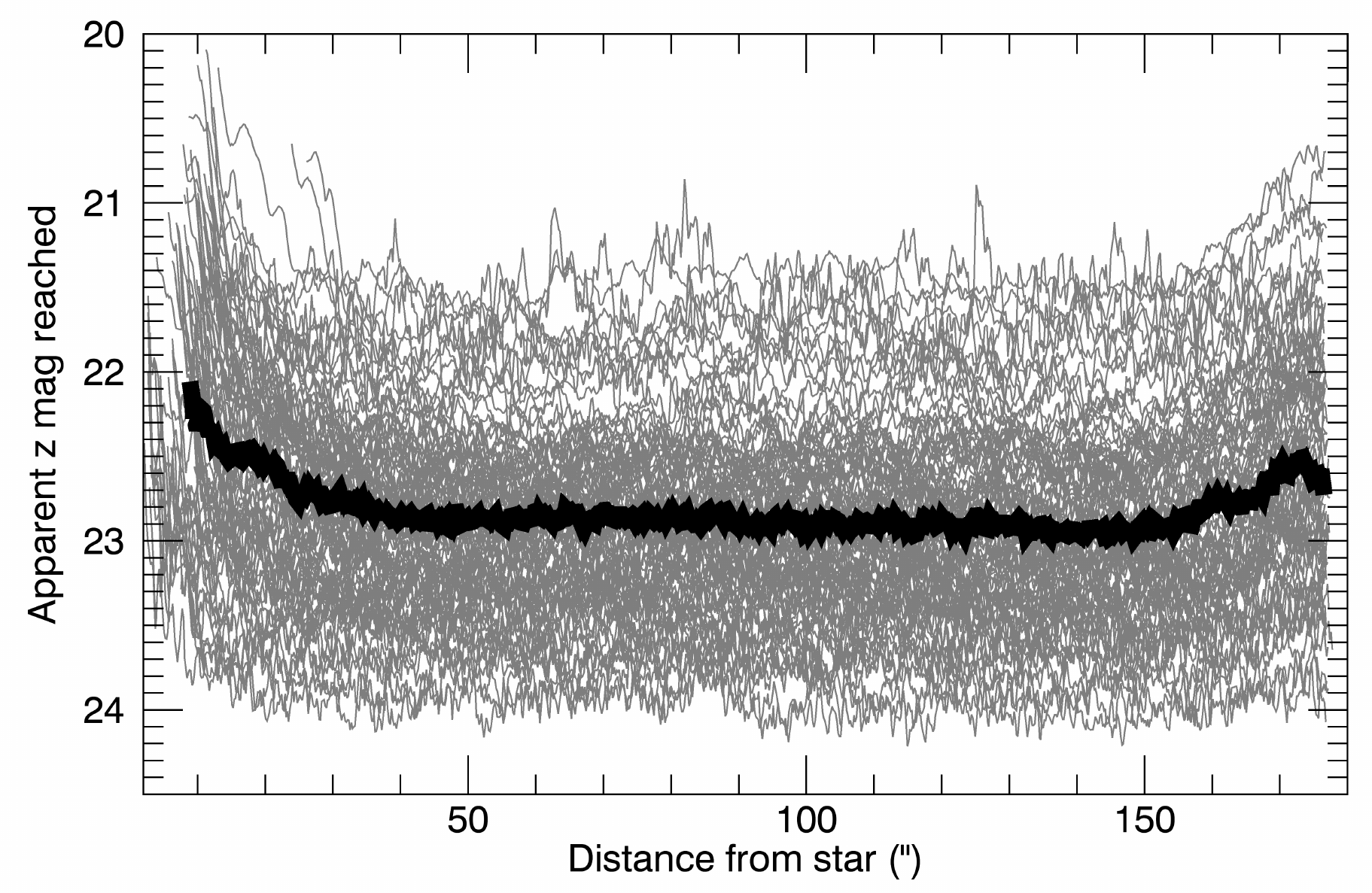}
\includegraphics[width=8.8cm]{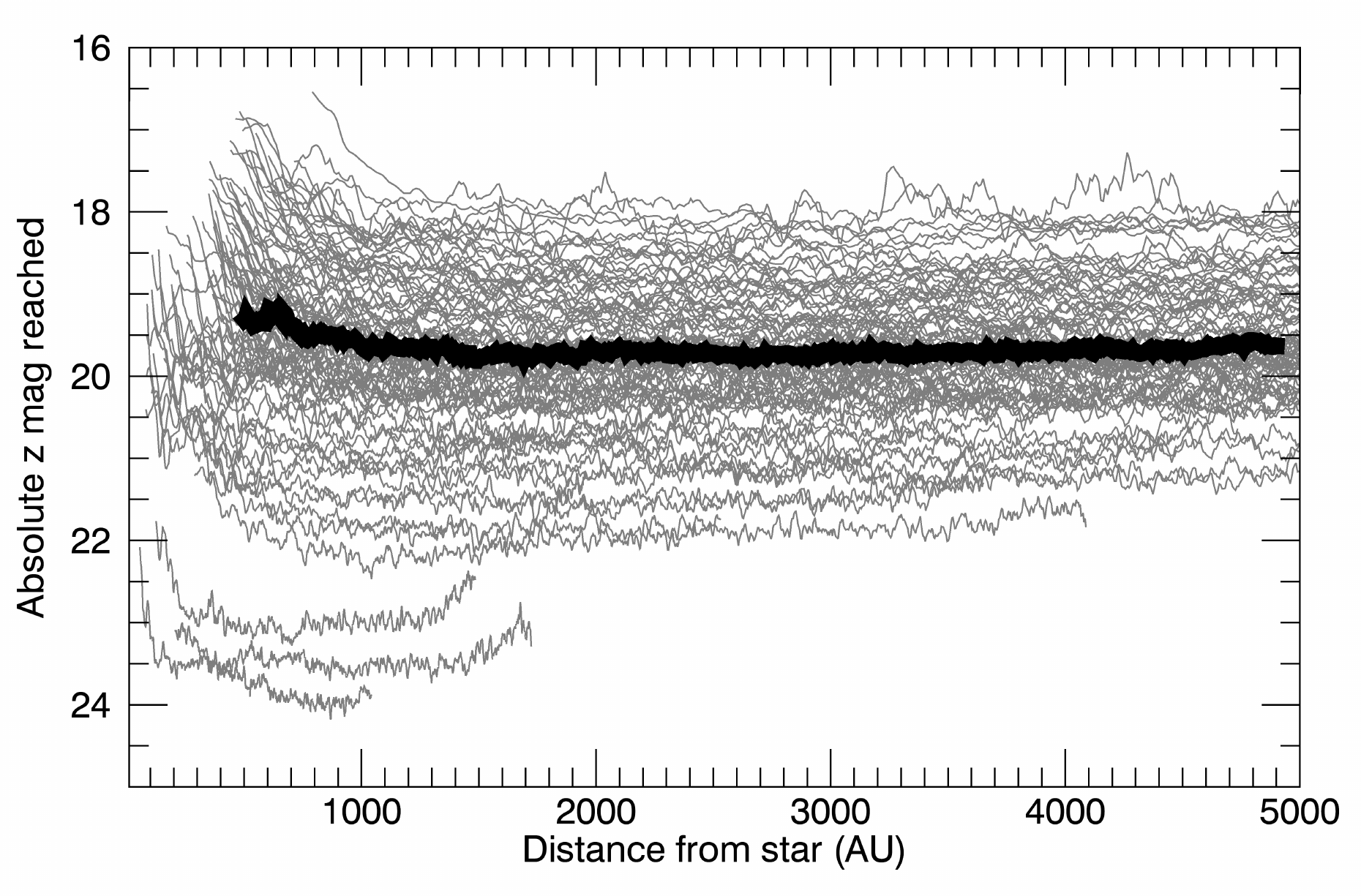}
%ccurves_all_appmag_asecpng.pdf}
%ccurves_all_absmag_AUpng.pdf}%png.pdf}
\caption{Left: apparent magnitude limit (5 $\sigma$) as a function of angular separation for all stars in the sample. The median apparent magnitude on the plateau is \z=22.9. Right: corresponding absolute magnitudes and projected physical separations in au, computed with a distance equal to the mean of the ranges listed in Table~2. The median curves are plotted in black. }
\label{fig:contrast}
\end{center}
\end{figure*}

The 5$\sigma$ detection limits based on background brightness were evaluated for every median-combined \z-magnitude image as a function of angular separation. At each angular separation step, this value is the standard deviation of the sky-subtracted flux in 180 circular apertures (1 FWHM radii), at this distance, located all around the target. The flux in the sky was evaluated for each aperture using an annulus with a radius 2 and 3 times the FWHM. This yielded an upper limit on the flux that a companion could have without being detected at 5$\sigma$. The limiting magnitude is fainter at further separations from the star. A plateau is typically reached at an angular separation of 20\arcsec\ and lasts up to the limits of the field, at an angular separation of \env 155\arcsec. The detection limits are shown in Figure~\ref{fig:contrast} and in Table~5. Average distances, corresponding to the centers of the ranges given in Table~2, were used to convert the angular separations to physical separations in astronomical units in the right panel of Figure~\ref{fig:contrast} and in Table~5. For clarity of presentation, these central values were also used to convert the apparent magnitude to absolute magnitude in Figure~\ref{fig:contrast}, while the full distance ranges were used to calculate the absolute magnitude ranges given in Table~5. For the most distant stars, the plateau where the survey is the most sensitive is not reached before a projected distance of 1000\,au or more and extends to separations that are well above 5000\,au.

The masses corresponding to limiting magnitudes can then be computed from the age of each star using the substellar hot-start evolutionary models of \citet{Baraffe2003}\footnote{Available at \url{http://phoenix.ens-lyon.fr/Grids/BT-Settl/CIFIST2011/ISOCHRONES/}.}. The full ranges of absolute magnitudes, distances, and ages were used to assess the limiting mass ranges (see Table~5). The \z\ apparent magnitudes in the 21.5--23.9 range were reached on the plateau, with a median value of \z=22.9. Considering the average of the lower and upper values for the distance and age ranges listed in Table~2, this corresponds to masses in the range 5--12\,\MJ.

% changé 3 pour 5 sigma! changé les notes en bas de page pour inclure la mention que le range donné de séparation en AU est calculé avec la dsitance moyenne
%tout ajuster pour ajouter une colonne vide entre range of sep et mag limit (pour que les clines ne se touchent pas
%\rm{a} pour les tablenote en bas de la table (pour pas que ce soit en italique)
%added ``full'' range and ``full'' ranges in notes b and c
%remplacé \asec par \arcsec
\begin{deluxetable*}{hllllccccc}
\renewcommand\thetable{5}
\tablecaption{5$\sigma$ Detection limits}
\label{tab-sensitivity}
\tablehead{
\nocolhead{i_s} & \colhead{2MASS} & \multicolumn{4}{c} {Range of separation} && \multicolumn{2}{c} {Magnitude limit} & \colhead{Mass limit\tablenotemark{c}} \\
\cline{3-6}
\cline{8-9}
\nocolhead{} & \colhead{designation} & \colhead{min} & \colhead{max} & \colhead{min\tablenotemark{a}} & \colhead{max\tablenotemark{a}} && \colhead{Apparent $z$} & \colhead{Absolute $z$\tablenotemark{b}}  & \colhead{}\\
\nocolhead{} & \colhead{} & \colhead{(\arcsec)} & \colhead{(\arcsec)} & \colhead{(au)} & \colhead{(au)}  && \colhead{} & \colhead{}  & \colhead{(\MJ)}}
\startdata
 0 & J00040288-6410358 &  4 & 150 &  208 &  7010 && 23.1 & 19.6--19.9 &  4.9-- 5.5\\
 1 & J00172353-6645124 & 32 & 152 & 1284 &  5969 && 22.8 & 19.7--20.0 &  4.5-- 4.9\\
 2 & J00325584-4405058 &  5 & 152 &  249 &  7040 && 23.0 & 19.1--20.6 &  3.6--11.7\\
 3 & J00374306-5846229 &  2 & 128 &  122 &  6371 && 22.8 & 18.9--19.9 &  3.1--12.2\\
 4 & J01071194-1935359 & 43 & 158 & 1786 &  6518 && 22.8 & 18.6--22.3 &  3.2--13.1\\
 5 & J01123504+1703557 & 20 & 155 &  950 &  7296 && 23.3 & 19.8--20.0 &  7.3-- 9.4\\
 6 & J01132958-0738088 & 27 & 123 & 1360 &  6055 && 22.6 & 18.7--19.6 &  3.2--28.9\\
 7 & J01220441-3337036 & 32 & 144 & 1262 &  5647 && 21.8 & 18.7--18.9 &  6.0-- 6.8\\
 8 & J01351393-0712517 & 26 & 164 &  999 &  6229 && 23.3 & 20.2--20.5 &  3.4-- 4.5\\
 9 & J01415823-4633574 &  5 & 155 &  223 &  6265 && 22.9 & 19.7--20.0 &  4.8-- 5.4\\
10 & J01484087-4830519 & 30 & 166 & 1094 &  5993 && 22.8 & 19.9--20.1 &  7.1-- 9.3\\
11 & J01521830-5950168 & 27 & 153 & 1087 &  5977 && 22.7 & 19.6--19.8 &  5.0-- 5.5\\
12 & J02045317-5346162 & 10 & 104 &  424 &  4301 && 22.9 & 19.7--19.9 &  4.9-- 5.4\\
13 & J02070176-4406380 & 23 & 161 & 1020 &  6960 && 23.3 & 20.0--20.2 &  4.3-- 4.9\\
14 & J02155892-0929121 & 29 & 119 & 1252 &  5139 && 22.5 & 19.2--19.4 &  5.6-- 6.0\\
15 & J02215494-5412054 &  3 & 148 &  148 &  5794 && 23.4 & 20.2--20.6 &  3.7-- 4.5\\
16 & J02224418-6022476 & 26 & 156 &  816 &  4849 && 22.8 & 20.2--20.4 &  3.9-- 4.6\\
17 & J02251947-5837295 & 13 & 136 &  567 &  5803 && 22.0 & 18.7--19.0 &  6.0-- 6.7\\
18 & J02303239-4342232 & 36 & 157 & 1907 &  8196 && 23.3 & 19.6--19.8 &  5.1-- 5.5\\
19 & J02340093-6442068 &  3 & 148 &  151 &  6823 && 22.4 & 18.9--19.2 &  5.7-- 6.3\\
20 & J02485260-3404246 & 29 & 158 & 1271 &  6820 && 23.1 & 19.8--20.1 &  4.6-- 5.3\\
21 & J02564708-6343027 & 19 & 151 & 1074 &  8317 && 22.0 & 18.2--18.6 &  6.3-- 8.0\\
22 & J03050976-3725058 & 32 & 154 & 2348 & 11156 && 22.5 & 18.1--18.3 &  6.8-- 8.1\\
23 & J03350208+2342356 &  6 & 155 &  267 &  6597 && 23.1 & 19.9--20.1 &  4.4-- 4.7\\
24 & J03494535-6730350 & 21 & 154 & 1766 & 12524 && 23.3 & 18.7--18.9 &  6.0-- 6.8\\
25 & J04082685-7844471 & 19 &  98 & 1038 &  5335 && 22.7 & 19.0--19.0 &  5.8-- 6.7\\
26 & J04091413-4008019 & 24 & 140 & 1574 &  8842 && 22.6 & 18.4--18.7 &  6.1-- 7.3\\
27 & J04213904-7233562 & 28 & 160 & 1524 &  8525 && 22.6 & 18.8--19.2 &  5.8-- 6.5\\
28 & J04240094-5512223 & 36 & 158 & 2472 & 10619 && 23.2 & 18.9--19.2 &  5.7-- 6.4\\
29 & J04363294-7851021 & 15 & 121 &  885 &  6810 && 22.3 & 18.4--18.8 & 10.6--14.0\\
30 & J04365738-1613065 & 33 & 150 &  760 &  3457 && 22.5 & 19.8--22.1 &  4.6-- 5.3\\
31 & J04402325-0530082 & 12 & 144 &  119 &  1412 && 23.4 & 23.5--23.5 & $<$ 3.2\\
32 & J04433761+0002051 & 15 & 156 &  382 &  3988 && 23.1 & 20.8--21.4 &  2.2-- 2.7\\
33 & J04440099-6624036 & 14 & 177 &  797 &  9573 && 22.7 & 18.9--19.2 &  5.7-- 6.4\\
34 & J04480066-5041255 & 37 & 148 & 1975 &  7733 && 22.5 & 18.8--19.1 &  5.8-- 6.6\\
35 & J04533054-5551318 & 58 & 148 &  652 &  1650 && 22.4 & 22.1--22.2 &  2.5-- 3.4\\
36 & J04571728-0621564 & 22 & 107 & 1020 &  4848 && 22.8 & 19.4--19.7 &  8.0--10.7\\
37 & J04593483+0147007 & 43 & 155 & 1121 &  4035 && 22.3 & 20.1--20.4 &  3.8-- 4.5\\
38 & J05090356-4209199 & 24 & 174 &  916 &  6450 && 22.8 & 19.1--21.4 &  2.3-- 6.2\\
39 & J05100427-2340407 & 42 & 155 & 2105 &  7612 && 22.5 & 18.9--19.3 &  5.5-- 6.4\\
40 & J05142878-1514546 & 10 & 172 &  657 & 10377 && 23.7 & 19.6--20.1 &  4.6-- 5.5\\
41 & J05241317-2104427 & 16 & 163 &  824 &  8326 && 23.7 & 20.0--20.4 &  3.9-- 5.0\\
42 & J05241914-1601153 & 37 & 168 &  705 &  3192 && 22.3 & 20.4--21.6 &  2.6-- 4.0\\
43 & J05254166-0909123 & 32 & 160 &  678 &  3323 && 23.1 & 21.3--21.8 &  3.2-- 5.5\\
44 & J05332558-5117131 & 33 & 141 & 1730 &  7378 && 22.9 & 19.1--19.5 &  5.5-- 6.1\\
45 & J05335981-0221325 & 29 & 144 & 1009 &  4924 && 22.3 & 19.4--19.9 &  4.5-- 5.2\\
46 & J05392505-4245211 & 25 & 161 & 1195 &  7498 && 23.2 & 19.5--20.4 &  3.9-- 5.7\\
47 & J05395494-1307598 & 11 & 160 &  804 & 10907 && 23.4 & 19.0--19.5 &  5.3-- 6.2\\
48 & J05470650-3210413 & 16 & 157 &  860 &  8209 && 22.9 & 19.1--19.7 &  5.2-- 6.1\\
49 & J05575096-1359503 & 21 & 153 &  864 &  6110 && 23.4 & 19.9--20.9 &  3.1--11.5\\
50 & J06045215-3433360 & 31 & 146 &  265 &  1229 && 22.6 & 23.0--23.0 & $<$ 2.3\\
51 & J06085283-2753583 &  5 & 160 &  159 &  4305 && 23.1 & 20.5--21.5 & $<$ 3.1\\
52 & J06112997-7213388 & 12 & 153 &  606 &  7216 && 21.9 & 18.5--18.6 &  6.2-- 7.7\\
53 & J06131330-2742054 & 32 & 160 &  940 &  4726 && 23.5 & 21.1--21.3 &  2.2-- 2.6\\
54 & J06434532-6424396 & 24 & 176 & 1310 &  9546 && 23.5 & 19.6--20.0 &  4.7-- 5.7\\
55 & J08173943-8243298 & 39 & 156 & 1071 &  4232 && 22.9 & 20.6--20.9 &  2.3-- 3.1\\
56 & J08471906-5717547 & 16 & 166 &  368 &  3670 && 22.2 & 20.3--20.7 &  5.9-- 8.3\\
57 & J10260210-4105537 & 26 & 161 & 1645 &  9826 && 23.4 & 19.3--19.7 &  3.7-- 4.4\\
58 & J10285555+0050275 & 71 & 147 &  505 &  1041 && 23.2 & 23.9--23.9 & $<$ 2.3\\
59 & J11115267-4401538 & 12 & 156 &  434 &  5333 && 23.4 & 20.3--21.2 &  3.7-- 7.3\\
60 & J11305355-4628251 &  7 & 160 &  435 &  9929 && 23.7 & 19.3--20.2 &  4.2-- 9.1\\
61 & J11592786-4510192 & 15 & 176 &  828 &  9723 && 23.9 & 19.8--20.7 &  3.1-- 4.0\\
62 & J12210499-7116493 & 23 & 157 & 2288 & 15464 && 23.6 & 18.4--18.9 &  3.3-- 5.5\\
63 & J12265135-3316124 & 15 & 175 & 1054 & 11594 && 23.6 & 19.4--19.6 &  3.7-- 4.4\\
64 & J12300521-4402359 & 11 & 177 &  809 & 12220 && 23.9 & 19.3--20.2 &  3.5-- 4.4\\
65 & J12383713-2703348 & 39 & 177 &  914 &  4087 && 23.6 & 21.7--21.9 &  2.9-- 4.2\\
66 & J14284804-7430205 & 19 & 171 &  909 &  7896 && 23.5 & 19.3--21.6 &  5.0--24.2\\
67 & J14361471-7654534 & 17 & 173 &  629 &  6071 && 23.9 & 20.7--21.8 &  2.6--16.2\\
68 & J15244849-4929473 & 13 & 175 &  320 &  4222 && 21.8 & 19.8--20.0 &  7.4-- 9.6\\
69 & J15310958-3504571 & 12 & 171 &  855 & 12031 && 23.2 & 18.6--19.5 &  3.4-- 5.1\\
70 & J16430128-1754274 & 12 & 161 &  516 &  6620 && 23.2 & 19.7--20.7 &  2.5--10.0\\
71 & J16572029-5343316 & 12 & 176 &  670 &  9188 && 21.6 & 17.9--18.1 &  6.2-- 7.1\\
72 & J18420694-5554254 & 19 & 157 & 1007 &  8330 && 21.7 & 17.9--18.2 &  6.1-- 7.0\\
73 & J19225071-6310581 & 23 & 150 & 1329 &  8645 && 22.0 & 17.9--18.5 &  5.8-- 9.0\\
74 & J19355595-2846343 &  4 & 175 &  134 &  5445 && 22.7 & 19.8--20.8 &  3.1-- 9.4\\
75 & J19560294-3207186 & 42 & 159 & 2450 &  9250 && 22.5 & 18.5--18.8 &  5.4-- 6.0\\
76 & J20004841-7523070 & 12 & 150 &  399 &  4833 && 22.3 & 19.6--20.0 &  4.4-- 5.0\\
77 & J20013718-3313139 & 38 & 168 & 2375 & 10464 && 22.9 & 18.8--19.1 &  5.2-- 5.7\\
78 & J20100002-2801410 & 25 & 161 & 1236 &  7761 && 23.0 & 19.5--19.7 &  4.7-- 5.1\\
79 & J20333759-2556521 & 15 & 169 &  758 &  8182 && 22.9 & 19.4--19.7 &  4.7-- 5.2\\
80 & J20465795-0259320 & 26 & 161 & 1236 &  7430 && 23.0 & 19.6--19.8 &  7.8--10.1\\
81 & J21100535-1919573 & 45 & 170 & 1491 &  5629 && 23.5 & 20.7--21.0 &  2.1-- 2.8\\
82 & J21265040-8140293 &  3 & 148 &  103 &  4757 && 22.6 & 19.9--20.2 &  3.1-- 9.4\\
83 & J21471964-4803166 & 13 & 153 &  733 &  8453 && 22.9 & 18.7--19.9 &  4.6--12.8\\
84 & J21521039+0537356 & 31 & 156 &  960 &  4768 && 22.7 & 19.9--20.7 &  5.9-- 9.3\\
85 & J22021626-4210329 & 35 & 164 & 1632 &  7554 && 23.0 & 19.6--19.9 &  5.0-- 5.5\\
86 & J22440873-5413183 & 20 & 137 &  979 &  6587 && 21.7 & 18.1--18.4 &  6.8-- 8.0\\
87 & J22470872-6920447 & 14 & 155 &  797 &  8531 && 22.5 & 18.7--18.9 & 10.2--13.0\\
88 & J23131671-4933154 & 18 & 113 &  731 &  4556 && 22.6 & 19.5--19.7 &  5.3-- 5.7\\
89 & J23221088-0301417 & 35 & 155 & 1361 &  5891 && 22.8 & 19.5--20.4 &  3.6--23.0\\
90 & J23285763-6802338 & 20 & 150 &  963 &  7211 && 22.8 & 19.2--19.5 &  5.5-- 6.0\\
91 & J23301341-2023271 & 46 & 154 &  760 &  2501 && 23.0 & 21.8--22.1 &  2.4-- 2.5\\
92 & J23320018-3917368 & 35 & 135 &  818 &  3127 && 23.3 & 21.4--21.6 &  3.6-- 5.2\\
93 & J23452225-7126505 & 14 & 154 &  652 &  6930 && 21.5 & 18.1--18.4 &  6.8-- 8.2\\
94 & J23474694-6517249 & 21 & 150 &  986 &  6904 && 23.1 & 19.7--19.9 &  5.0-- 5.4\\
\enddata
\tablenotetext{\rm{a}}{Considering the average of the distance range given in Table \ref{tab-staragedist}} 
\tablenotetext{\rm{b}}{Considering the full distance range given in Table \ref{tab-staragedist}} 
\tablenotetext{\rm{c}}{Using the full distance and age ranges given in Table \ref{tab-staragedist} in the \citet{Baraffe2003} evolutionary models.}
\end{deluxetable*}

\begin{figure}[htbp]
\begin{center}
\includegraphics[width=8.5cm]{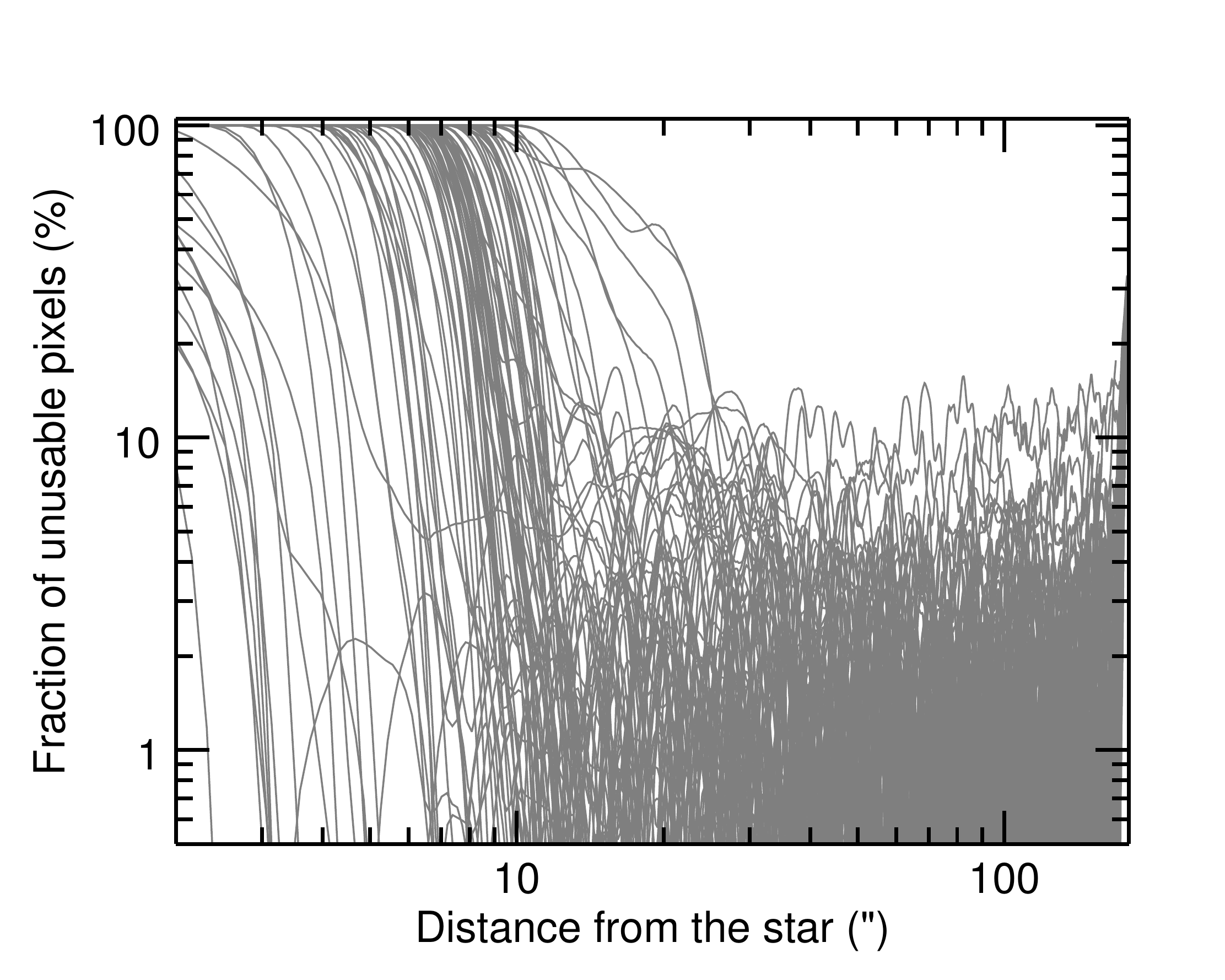}%Figures/usable_asec_xlog.eps}
\caption{Fraction $1-f_{u}$ of pixels where a companion cannot be found, considering bad pixels and background stars in the field. Beyond \env10\arcsec, $>$98\% of putative companions would have been identified for the large majority of stars. A few low-galactic-latitude stars have a lower plateau value.}
\label{fig:usable}
\end{center}
\end{figure}

For each target, it is possible to compute the fraction $f_{u}$ of \z\ image pixels where a companion could have been detected at 5$\sigma$. This takes into account the bad pixels and background sources that hinder the detection of a companion. This quantity is represented in Figure~\ref{fig:usable} as a function of separation for all sample stars. It shows that beyond 10\arcsec, typically more than 98\% of putative companions should have been detected. 
For the stars that are the closest to the galactic plane, the density of objects is higher, and the fraction of objects that can be recovered can be lower (down to $96\%$). This is taken into account in the computation of completeness limits in Section \ref{complete}. 

\subsection{Completeness Maps and Survey sensitivity} 
\label{complete}
The detection limits in terms of absolute magnitudes and projected separations determined in Section \ref{contrast} can be used to evaluate the sensitivity of the survey to planets of a given mass and semimajor axis. The method used here is similar to that described by \citet{Nielsen:2008uu}. 

\begin{figure}[htbp]
\begin{center}
\includegraphics[width=8.8cm]{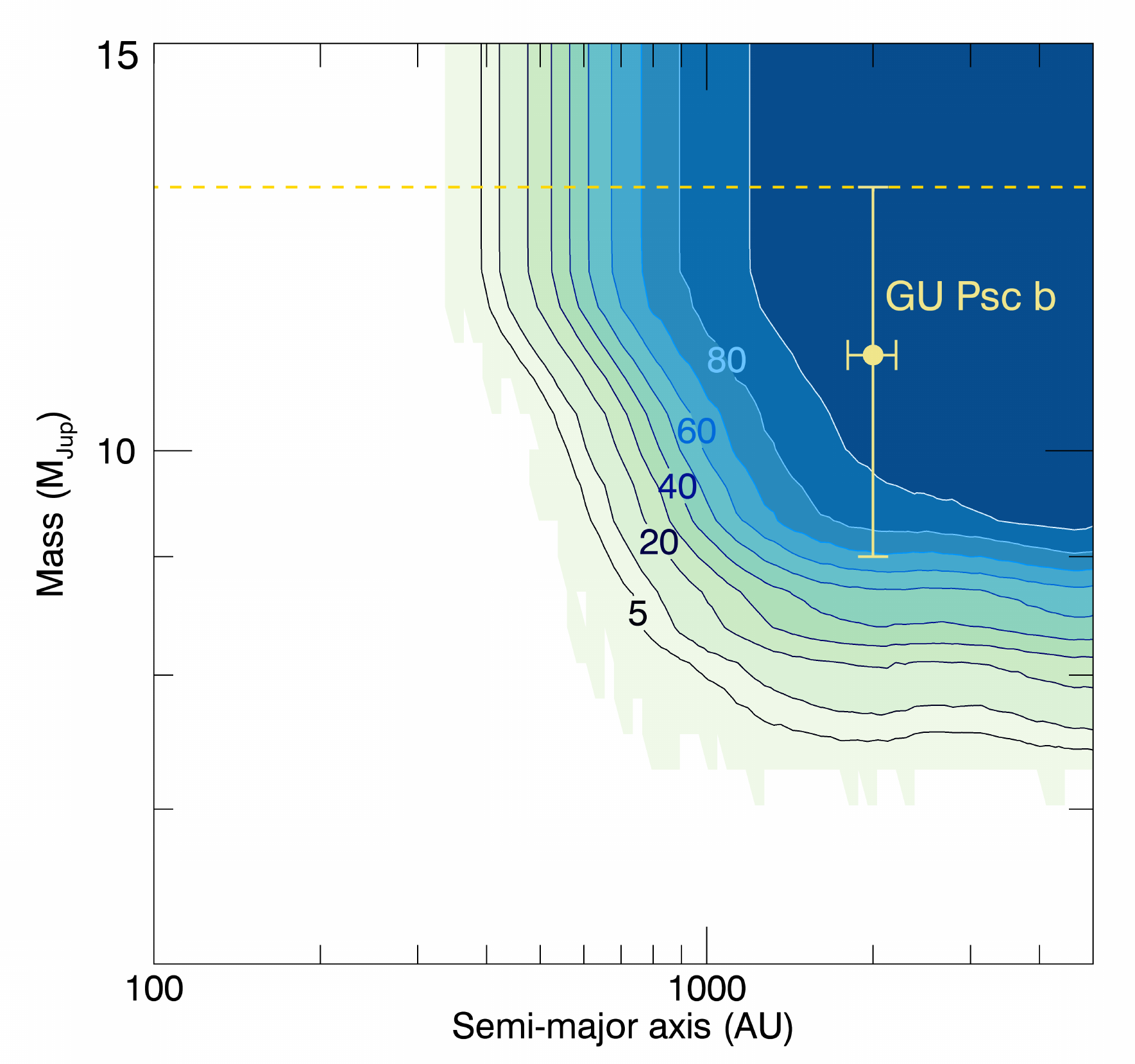}%Figures/completeness_map_Nielsen_star_5sig5.pdf}
\caption{Completeness map for the star GU Psc. The contours indicate the fraction of planets that would be recovered in percent, considering a uniform eccentricity distribution between 0 and 0.8, a random inclination and time of observation, the distance and age ranges given in Table~2, and the hot-start models of \citet{Baraffe2003}. The horizontal dashed line is the 13\,\MJ\ separation between planetary-mass objects and brown dwarfs. }%The map is cut above 15\,\MJ, since the survey is not complete for more massive brown dwarfs.
\label{fig:completenessGU}
\end{center}
\end{figure}

A Monte Carlo simulation was first used to build a completeness map for each star, that is, to assess what fraction of planets of a given mass and semimajor axis can be retrieved around it, considering the distribution of possible orbital parameters and considering its credible age and distance ranges. A grid of 100 $\times$ 100 masses and semimajor axes was built, spread uniformly in log space, for masses between 3 and 100\,\MJ\ and semimajor axes of 100 to 5000\,au. At each point of the grid, a population of 10,000 planets was simulated. The method described in \citet{Brandeker2006} and \citet{Brandt2014} was used to determine the distribution of projected separations in astronomical units from the semimajor axes, given a distribution of eccentricity and assuming a random viewing angle and time of observation. The eccentricity distribution function adopted here is that of \citet{Cumming2008}, that is, a uniform distribution between 0 and 0.8. The distance and age, sampled linearly within the ranges listed in Table~2, are used to convert physical projected separations to angular projected separations and to convert masses to absolute \z\ fluxes, using the evolutionary models of \citet{Baraffe2003}. The 5$\sigma$ detection curves computed in Section \ref{contrast} (Figure~\ref{fig:contrast}) can then be used to determine whether or not a given simulated planet would be bright enough to be recovered around its host. If so, the fraction of pixels where a companion can be found $f_{u}$ is taken as the detection probability. Repeating these steps for each simulated planet allows us to determine the fraction of planets that would have been detected around a star at each grid point. The resulting map is shown in Figure~\ref{fig:completenessGU} for GU Psc. 

\begin{figure*}[htbp]
\begin{center}
\includegraphics[trim=0cm 0.0cm 0cm 0cm, clip=true, height=7.5cm]{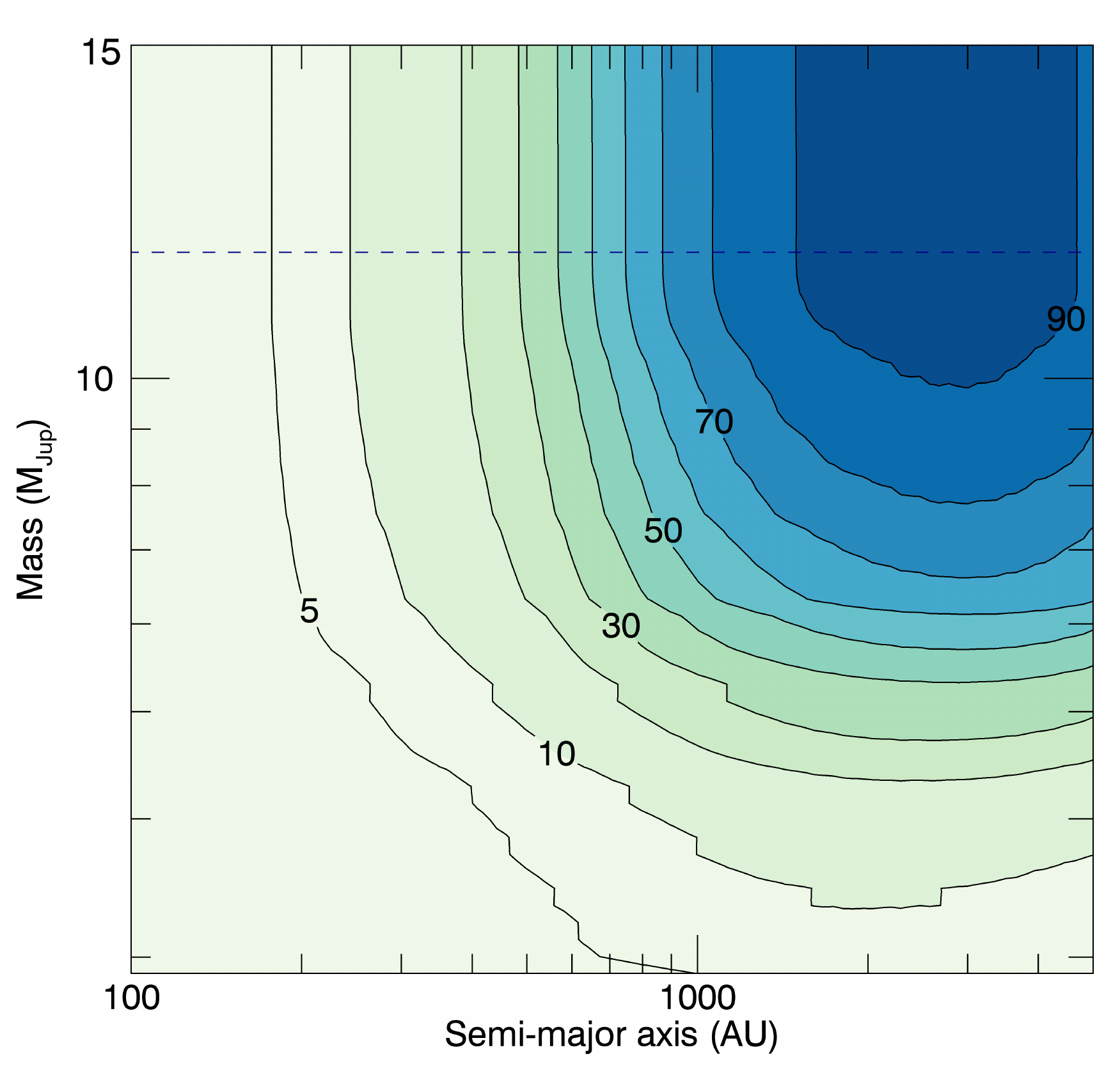}
\includegraphics[trim=0cm 0.5cm 0cm 0cm, clip=true, height=7.8cm]{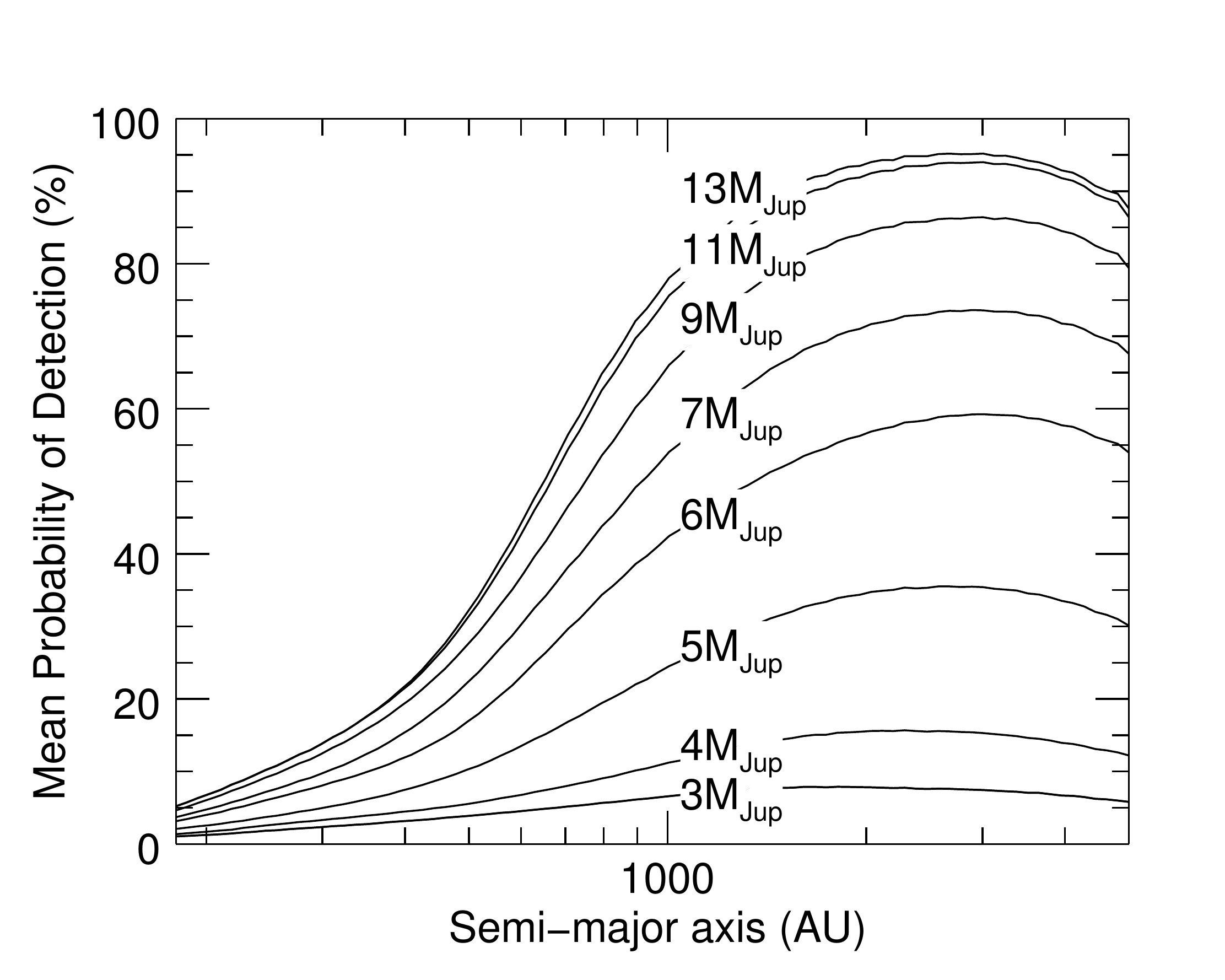}
\caption{Completeness map and mean detection probability for the survey. The left panel gives the mean detection probability in percentage with respect to mass and semimajor axis.  The horizontal dashed line is the 13\,\MJ\ separation between planetary-mass objects and brown dwarfs. The right panel shows the mean probability of detection vs. semimajor axis, for specific values of companion mass.} 
%completeness_map_Nielsen_5sig-1croppng.pdf}%completeness_map_Nielsen_10sig.eps}
%Figures/mean_prob_detection_5sig.eps
\label{fig:detlim}
\end{center}
\end{figure*}

Taking the sum of the maps for all stars allows us to assess the mean sensitivity for the entire survey (Figure~\ref{fig:detlim}), in terms of the fraction of stars in the survey for which a planet of a given mass and semimajor axis would have been detected. The figure demonstrates that the survey is most sensitive above 1000\,au, with a peak between 2000 and 4000\,au. The maximal detection probabilities are of 8\%, 36\%, 86\%, 94\%, and 95\% for masses of 3, 5,  9, 11, and 13\,\MJ, respectively. The survey is particularly sensitive to planets at the massive end of the planetary-mass range. The mean detection probabilities for 3\,\MJ\ companions are below 10\% for all semimajor axes. At separations of \env500\,au, the detection probabilities are nonnegligible: 10\% for 5\,\MJ\ and 30\% for 11\,\MJ. The probability of finding a planet at 2000\,au with the mass of GU~Psc~b (\env11\,\MJ) is over 90\%. At 100--200\,au, where most AO imaging surveys are most sensitive, the present survey has a small detection probability of less than 5\%, even for the most massive planets. %ca donne aussi le nombre de planètes qu'on aurait trouve s'il y avait eu une planète de cette masse et cette sep autour de toutes les etoiles

\subsection{Planet Frequency}
\label{subsec:frequency}
Using the results presented in Section \ref{complete} and the statistical formalism presented in \citet{Lafreniere2007}, it is possible to determine a credible interval for the fraction $f$ of late spectral type (K5--L5) stars that have at least one companion in a given mass and semimajor axis range. If the $N=95$ sample stars are enumerated $j=1 \dots N$, the results of this survey are summarized by the set $\{d_{j}\}$, where the value of $d$ is 1 for stars with a detected companion or 0 otherwise. The resulting set $\{d_j\}$ depends on the true fraction of stars $f$ that host a planet in the surveyed range of semimajor axes and masses. It is given by the binomial likelihood: 

\begin{equation}
\mathcal{L}(\{d_{j}\}|f)=\prod_{j=1}^{ N}(1-fp_{j})^{(1-d_{j})}(fp_{j})^{d_{j}}
\end{equation}

The completeness maps (as shown for GU~Psc~b in Figure~\ref{fig:completenessGU}) are used to determine $p_{j}$, which represents the probability of detecting a companion with a mass in a given range $[m_{min},m_{max}]$ and a semimajor axis in a given range $[a_{min},a_{max}]$. For each star, $p_{j}$ is taken to be the mean of the recovered planet fraction in all grid points for the mass and semimajor axis ranges considered. Since the grid is uniform in log mass and log $a$, this is equivalent to assuming log-uniform distributions for these two parameters. Bayes' theorem states that the posterior distribution, which is the probability density function of $f$ considering the results of the survey $\{d_{j}\}$, is given by

\begin{equation}
P(f|\{d_{j}\})=\frac{\mathcal{L}(\{d_{j}\}|f)P(f)}{\int_{0}^{1}{\mathcal{L}(\{d_{j}\}|f)P(f)\mathrm{d}f}}. 
\end{equation}
The denominator can be referred to as the \textit{marginalized likelihood}. The prior distribution $P(f)$ represents the best knowledge on the probability density for $f$ using only information independent from the current survey. In several direct imaging survey analyses, a flat prior distribution $P(f)=1$ was used. While simpler, a uniform prior is in general not mathematically equivalent to having no prior knowledge on the parameters. As an illustration of this concept, a change of coordinates can result in a different answer if a flat prior is used in both coordinate systems, and therefore the resulting posterior does not only depends on the likelihood model and the available data, but also depends on the way that the problem is parameterized. Applying Bayesian statistics in a way that only depends on the available data and the likelihood model requires using non informative priors (e.g., see \citealt{Berger2009}), which do not always correspond to flat priors. In a case with only one parameter, the non-informative prior can be derived in a simple way and is called Jeffrey's prior (see \citealt{Jeffreys1998}). The Jeffrey’s prior that is associated with the binomial likelihood is given by  

\begin{equation}
P(f)=\frac{1}{\pi}\frac{1}{\sqrt{f}}\frac{1}{\sqrt{1-f}}.
\end{equation}

\begin{figure*}[htbp]
\begin{center}
\includegraphics[width=15cm]{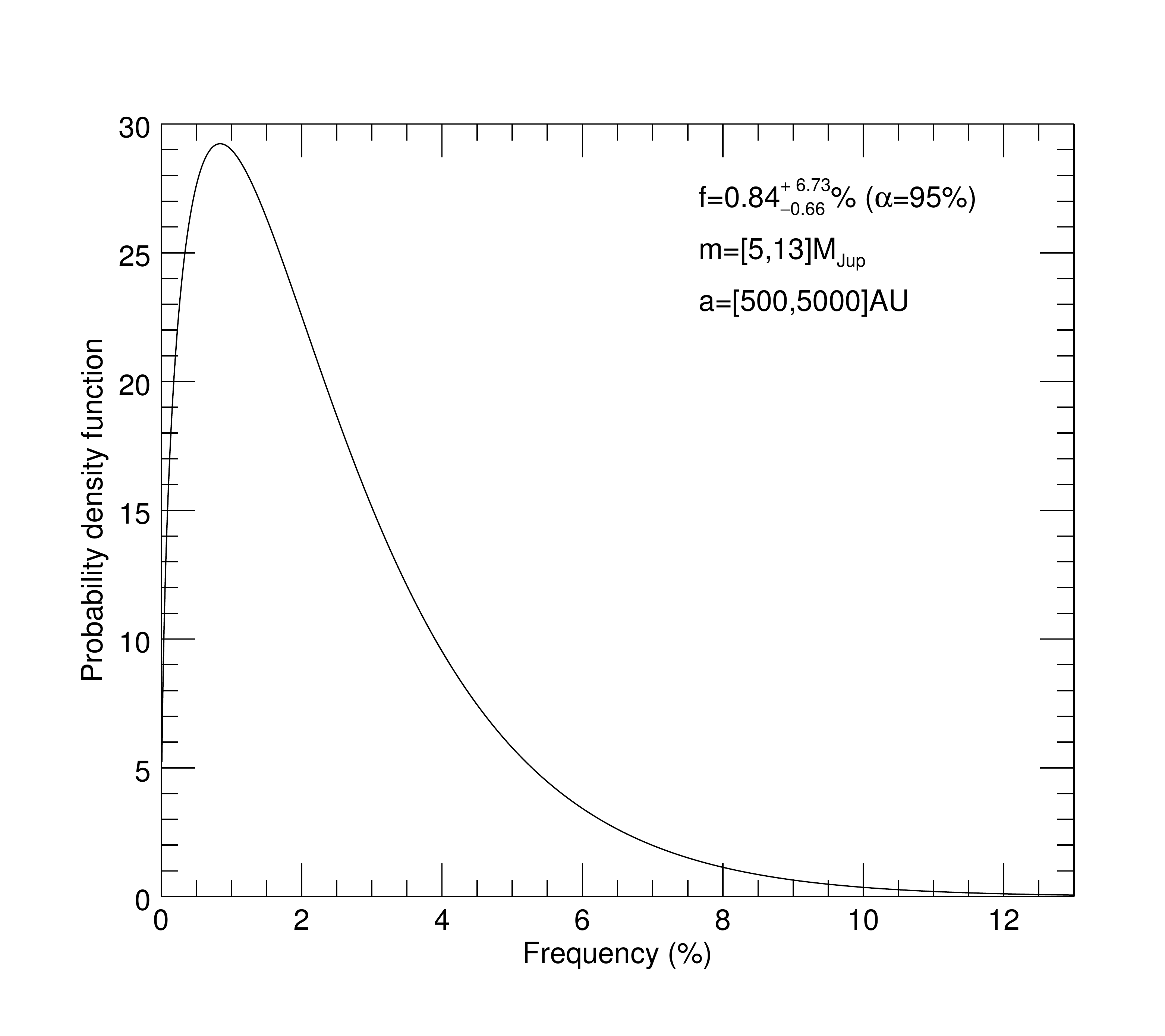}%pdf.eps}%completeness_map_Nielsen_10sig.eps}
\caption{Probability density function for the frequency of late spectral type (K5--L5) stars with at least a companion with masses in the range m $= [5,13]$\,\MJ\ and semimajor axes in the range a $= [500,5000]$\,au.}
\label{fig:pdf}
\end{center}
\end{figure*}

As shown in Figure~\ref{fig:detlim}, the survey is particularly sensitive for semimajor axes between 500 and 5000\,au and masses between 5 and 13\,\MJ. The posterior distribution was thus computed for these ranges and is shown in Figure~\ref{fig:pdf}. This accounts for the detection of a single companion (GU Psc b) in these intervals. Only the projected separation of the companion (2000\,au) is known, but considering the eccentricity distribution of \citet{Cumming2008} and the random viewing time and inclination as in Section \ref{complete}, it can be shown that the semimajor axis of the companion is unlikely to have a semimajor axis above 5000\,au. The peak of this posterior distribution corresponds to the most likely value of $f$. Given a level of confidence $\alpha$, an equal-tail credible interval $[f_{min},f_{max}]$ can be determined using 

\begin{align}
\frac{1-\alpha}{2}&=\int_{0}^{f_{min}}p(f|\{d_{j}\})df,\\ \frac{1+\alpha}{2}&=\int_{f_{max}}^{1}p(f|\{d_{j}\})df.
\end{align}

The fraction of late spectral type (K5--L5) stars that have at least one companion in this semimajor axis and mass ranges is $0.84_{-0.66}^{+6.73}\%$ ($\alpha$=95\%). Note that if a flat prior had been assumed, the planet frequency would have been artificially larger with a wider confidence interval ($1.66_{-1.27}^{+7.22}\%$).

%%%%%%%%%%%%%%%%%%%%%%%%%%%%%%%%%%%%%%%%%%%%%%%%%%%%%%%%%%%%%%%%%%%%%%%%%

\section{Discussion}
\label{discussion}
The sensitivity of the present survey to planetary-mass companions (5--13\,\MJ) is maximized between 500 and 5000\,au, much farther than typical AO-assisted imaging surveys around similar stars, which are sensitive to up to 1000\,au at best. The small overlap can be used to compare the limits on the occurrence of companions around young M stars, and to study this occurrence as a function of the separation from the central star.

The meta-analysis of \citet{Bowler2016} puts an upper limit of $<7.3$\% ($95\%$ confidence level) on the frequency of 5--13\,\MJ\ companions at separations between 100 and 1000\,au around M stars. An analysis similar to that presented in Section \ref{subsec:frequency} and based on the present survey data was carried for these ranges. Only an upper limit was determined because GU~Psc~b is in all likelihood outside this range of semimajor axes considering the method described in Section \ref{complete} (there is less than a 15\% chance that GU~Psc~b has a semimajor axis below 1000\,au with a projected separation of 2000\,au). An upper limit of $<11.1\%$ was found at the same confidence level. This is consistent with the \citet{Bowler2016} result. The \citet{Bowler2016} survey is more constraining because it includes more stars (119 compared to 95), but also because the present survey is only moderately sensitive in these ranges: the average detection probability for 13\,\MJ\ at 1000\,au is close to 80\% but only $25\%$ for 5\,\MJ. \citet{Lafreniere2007} derived an analytical expression for the planet frequency $f_{\rm{max}}$ in the special case of nondetections:

\begin{equation}
f_{\rm{max}}\sim\frac{-\ln{(1-\alpha)}}{N\langle p_{j}\rangle},
\end{equation}

where $\langle p_{j}\rangle$ is the average planet detection probability, $N$ the total number of stars in the survey, and $\alpha$ the confidence interval level. This approximation is valid for $N\langle p_{j}\rangle\gg1$. Since the same intervals were used in the present survey and the \citet{Bowler2016} analysis, both results can be combined, assuming $\alpha=0.95$, to derive an upper limit of $<4.4\%$ for the fraction of late spectral type (K5--L5) stars with at least a giant planetary-mass companion in the mass range [5,13]\,\MJ\ with semimajor axis $<$1000\,au.

\citet{Lannier2016} found that $2.3_{-0.7}^{+2.9}\%$ (1$\sigma$ confidence) of M stars have a 2--14\,\MJ\ companion between 8 and 400\,au. The present survey is not sensitive to companions between 2 and 5\,\MJ\ and below 400\,au, so it is not relevant to compute a frequency in this range of parameters. It is, however, interesting to note that the fraction obtained by \citet{Lannier2016} is similar to that found here for more massive and more distant planets. It is also similar to the $2_{-1}^{+3}\%$ frequency derived from radial velocity data \citep{Bonfils2011} for less massive giant planets (up to \env3\,\MJ) very close to low-mass older stars (periods between 10 and 100 days; main-sequence stars). All planet surveys so far have demonstrated that gas giants are rare beyond \env10\,au around low-mass stars, as expected from planet formation models. The survey presented here yielded a planet frequency similar to those found for closer-in planets within uncertainties, although it spans a much wider separation interval. The planet frequency thus seems to remain similar over three orders of magnitude in orbital separations, despite the fact that planets in these regimes likely form through different mechanisms.

There is no agreement at this stage as to whether planetary-mass companions at wide separations are correlated with the stellar mass, as suggested for closer-in companions \citep{Johnson2007,Borucki2011}. \citet{Lannier2016} find that such a correlation probably exists for substellar companions that have a low mass ratio ($Q<1\%$). This is in agreement with the conclusion of \citeauthor{Montet2014} (\citeyear{Montet2014}; from a combination of direct imaging and radial velocity) and \citeauthor{Clanton2014} (\citeyear{Clanton2014}; combination of microlensing and radial velocity), that giant planets are less frequent around low-mass stars. However, they do not find evidence for a correlation at higher mass ratio values ($1\%<Q<5\%$). GU~Psc~b, with $Q\sim3\%$, falls in that regime. In their meta-analysis, \citet{Bowler2016} and \citet{Galicher2016} do not find evidence that there are fewer giant planets around low-mass stars; in both surveys, the frequencies derived for host stars of different masses are compatible with each other. While the present survey confirms the existence -- albeit rare -- of planetary-mass companions at wide separations, more detections are required to determine whether the presence of these are correlated with stellar host mass. 

%%%%%%%%%%%%%%%%%%%%%%%%%%%%%%%%%%%%%%%%%%%%%%%%%%%%%%%%%%%%%%%%%%%%%%%%%
\section{Conclusion}
\label{conclusion}
The PSYM-WIDE survey allowed us to search for planetary-mass companions around 95 low-mass stars (spectral types K5--L5) that are members of young associations. It used Gemini GMOS \i\ and \z\ imaging to identify them via their distinctively red \iz\ color and allowed us to establish a frequency of stars with at least one companion of 0.84$_{-0.66}^{+6.73}$\% (95\% confidence) in the mass range 5--13\,\MJ\ and with semimajor axes range 500--5000\,au. 

The only planet discovered through this survey (GU~Psc~b; \citealp{Naud2014}) and other substellar companions discovered via direct imaging (e.g., the \env23\,\MJ\ brown dwarf HIP~78530 B; \citealp{Lafreniere2008, Lafreniere2010} or Ross~458 (AB) c, a distant planetary-mass companion to a M0.5+M7 binary, \citep{Burgasser2010_Ross,Goldman2010} are too widely separated from their stars for in situ formation by either core accretion or gravitational instability. This suggests that other mechanisms, such as direct formation through the turbulent fragmentation of a prestellar core \citep{Padoan2002, Bate2003} or ejection through interaction with a massive companion, could be at play in these cases.

As demonstrated by the in-depth photometric and spectroscopic study of GU Psc b \citep{Naud2014} and the study of its light curve evolution \citet{Naud2017b}, wide planetary-mass companions are amenable to a level of characterization that is useful in assessing the characteristics of closer-in giant planets, which are much harder to study. Further surveys to identify wide-separation exoplanets would be valuable, especially deeper ones that are focused on the identification of less-massive giant planets.  
New detections would contribute to investigating possible correlations with the mass of the host star, and more generally the various formation mechanisms at play. The WEIRD survey (Wide orbit Exoplanet search with InfraRed Direct imaging; \citealp{Baron2015}), an ongoing effort using Spitzer and ground-based facilities such as CFHT and Gemini, will provide better constraints on the presence of these very wide ($>$500--1000\,au) planetary-mass companions. The observations are obtained at 3.6 and 4.5\,\micron\ and are thus sensitive to planets down to about the mass of Saturn (0.3\,\MJ). 
 
 %%%%%%%%%%%%%%%%%%%%%%%%%%%%%%%%%%%%%%%%%%%%%%%%%%%%%%%%%%%%%%%%%%%%%%%%%
\subsection*{Acknowledgments}
The authors would like to thank Julien Rameau for his valuable suggestions and helpful discussions. They are also very grateful for the help of the Pan-STARRS1 and SkyMapper teams for providing data and the support for using it to do the photometric calibration of the data. They would also like to thank the anonymous referee for constructive comments and suggestions that improved the overall quality of the paper. This work was financially supported by the Natural Sciences and Engineering Research Council (NSERC) of Canada and the Fond de Recherche Qu\'{e}b\'{e}cois - Nature et Technologie (FRQNT; Qu\'{e}bec). 
This publication makes use of data products from the Two Micron All Sky Survey, which is a joint project of the University of Massachusetts and the Infrared Processing and Analysis Center, and funded by the National Aeronautics and Space Administration and the National Science Foundation, of the NASA Astrophysics Data System Bibliographic Services, the VizieR catalog access tool, and the SIMBAD database operated at CDS, Strasbourg, France. It also made use of the L and T dwarf data archive \url{http://staff.gemini.edu/~sleggett/LTdata.html}.

This work also used data from the Sloan Digital Sky Survey III (SDSS-III). Funding for this survey has been provided by the Alfred P. Sloan Foundation, the Participating Institutions, the National Science Foundation, and the U.S. Department of Energy Office of Science. The SDSS-III web site is \url{http://www.sdss3.org/}. SDSS-III is managed by the Astrophysical Research Consortium for the Participating Institutions of the SDSS-III Collaboration including the University of Arizona, the Brazilian Participation Group, Brookhaven National Laboratory, University of Cambridge, University of Florida, the French Participation Group, the German Participation Group, the Instituto de Astrofisica de Canarias, the Michigan State/Notre Dame/JINA Participation Group, Johns Hopkins University, Lawrence Berkeley National Laboratory, Max Planck Institute for Astrophysics, New Mexico State University, New York University, Ohio State University, Pennsylvania State University, University of Portsmouth, Princeton University, the Spanish Participation Group, University of Tokyo, University of Utah, Vanderbilt University, University of Virginia, University of Washington, and Yale University.

Data products from the Pan-STARRS were also used. PS1 has been made possible through contributions of the Institute for Astronomy, the University of Hawaii, the Pan-STARRS Project Office, the Max Planck Society and its participating institutes, the Max Planck Institute for Astronomy, Heidelberg, and the Max Planck Institute for Extraterrestrial Physics, Garching, The Johns Hopkins University, Durham University, the University of Edinburgh, Queen's University Belfast, the Harvard-Smithsonian Center for Astrophysics, the Las Cumbres Observatory Global Telescope Network Incorporated, the National Central University of Taiwan, the Space Telescope Science Institute, the National Aeronautics and Space Administration under Grant No. NNX08AR22G issued through the Planetary Science Division of the NASA Science Mission Directorate, the National Science Foundation under Grant No. AST-1238877, the University of Maryland, and Eotvos Lorand University (ELTE), and the Los Alamos National Laboratory.

Finally, SkyMapper data products were used. The national facility capability for SkyMapper has been funded through ARC LIEF grant LE130100104 from the Australian Research Council, awarded to the University of Sydney, the Australian National University, Swinburne University of Technology, the University of Queensland, the University of Western Australia, the University of Melbourne, Curtin University of Technology, Monash University, and the Australian Astronomical Observatory. SkyMapper is owned and operated by The Australian National University's Research School of Astronomy and Astrophysics. The survey data were processed and provided by the SkyMapper Team at ANU. The SkyMapper node of the All-Sky Virtual Observatory is hosted at the National Computational Infrastructure (NCI).

\end{document}